\documentclass[11pt,a4paper]{article}
\usepackage{jheppub1}
\usepackage{graphicx}
\usepackage{subcaption}
\usepackage{bm}
\usepackage{mathtools}
\usepackage{slashed}
\usepackage{physics}
\usepackage{hyperref}
\usepackage{verbatim}
\usepackage{tikz}
\usetikzlibrary{arrows.meta,calc}
\usepackage{xcolor}

\begin{document}
\title{Quantum Corrections to Page Curve of Charged Near-AdS$_2$ Black Holes}
\author[a]{Zi-Qing Xiao}
\emailAdd{xiaoziqing24@mails.ucas.ac.cn}
\affiliation[a]{International Centre for Theoretical Physics Asia-Pacific (ICTP-AP), University of Chinese Academy of Sciences (UCAS), Beijing, China.}
\abstract{We study the evaporation and Page curve of charged near-AdS$_2$ black holes coupled to a non-gravitating bath at fixed temperature and chemical potential. The low-energy dynamics is governed by the Schwarzian reparametrization mode together with a $U(1)$ phase mode. We average the boundary energy and the outgoing flux over these two soft modes and obtain corrected balance equations for the temperature and chemical potential. We then use the corrected background to calculate the no-island and island entropies and the Page time shifts. We find that the two soft sectors affect the Page transition in our low-temperature semiclassical regime. The $U(1)$ phase mode correction delays the Page transition, while the Schwarzian correction tends to move it earlier. The total Page time shift is therefore determined by the competition between the Schwarzian and $U(1)$ sectors. 
}
\maketitle
\noindent
\section{Introduction}

The Page curve is one of the basic probes of the black hole information problem. In
Hawking's semiclassical calculation, the radiation entropy keeps increasing during the
whole evaporation process. If the evaporation is unitary, this cannot be the final answer:
after the Page time the entropy should decrease. The island formula gives a way to obtain
such a unitary Page curve in many semiclassical models
\cite{Almheiri:2019psf,Almheiri:2019hni,Penington:2019npb,Almheiri:2020cfm,
Penington:2019kki,Almheiri:2019qdq,Almheiri:2019psy,Goto:2020wnk}.

Once the leading Page curve is understood, the next question is what happens when quantum effects change the evaporation law. The Page curve is evaluated on a time-dependent
black hole background. If quantum effects change the way the black hole loses energy and
charge, then the Page curve should also change. Near-extremal black holes provide a controlled setting for this question. The reason is that their low-energy dynamics is governed by near-AdS$_2$ physics, where the relevant
gravitational degrees of freedom reduce to boundary soft modes \cite{Almheiri:2014cka,Maldacena:2016upp,Mertens:2022irh}. In two dimensions there are
no propagating gravitational degrees of freedom in the bulk. The Schwarzian reparametrization mode describes the gravitational soft sector of the near-AdS$_2$ throat. For a charged black hole, the same low-energy theory also contains a compact $U(1)$ phase mode, which keeps track of the charge sector and the chemical potential. Thus the charged near-AdS$_2$ system gives a simple place where evaporation, charge relaxation, and quantum effects of the soft modes can be treated in one setup.

This point is supported by recent work on Schwarzian quantum effects. These effects appear
in the low-temperature thermodynamics of near-extremal black holes and in the logarithmic
corrections produced by near-horizon soft modes \cite{Iliesiu:2020qvm,Iliesiu:2022onk,Banerjee:2023quv,Kapec:2023ruw,Rakic:2023vhv,
Banerjee:2023gll,Maulik:2024dwq,Kapec:2024zdj,Kolanowski:2024zrq,
PandoZayas:2026vbg}. They also affect real-time observables, including Hawking emission,
absorption cross sections, and the evaporation history of near-extremal charged or rotating
black holes
\cite{Brown:2024ajk,Mohan:2024rtn,Maulik:2025hax,Emparan:2025sao,
Biggs:2025nzs,Lin:2025wof,Betzios:2025sct,Luo:2026epp,Rakic:2025svg}. Related
Schwarzian averages have also been used in response and transport problems.\footnote{
Examples include strange-metal transport, low-temperature fluid/gravity, shear viscosity,
decoherence, quasinormal modes, and shear correlators
\cite{Liu:2024gxr,Nian:2025oei,Cremonini:2025yqe,Li:2025vcm,Jiang:2025cyl,
Gouteraux:2025exs,Kanargias:2025vul}.}
These results suggest that the Schwarzian mode should not be viewed only as a correction
to the equilibrium partition function. It can also correct fluxes and real-time observables
\cite{Mertens:2019bvy}, and hence it can affect the evaporating background on which the
Page curve is evaluated.

There has also been progress on charged black holes with islands. Islands and Page curves
have been studied for Reissner--Nordstr\"om black holes, charged dilaton black holes,
one-sided charged black holes, and symmetry-resolved radiation entropies
\cite{Wang:2021woy,Kim:2021gzd,Yu:2021cgi,Qu:2024sby,Li:2023zgy}. These works show
that charge can affect the island saddle, the radiation entropy, and the Page transition.

A related question was raised recently from the thermodynamic side, where quantum
corrections to near-extremal Kerr-Newman black holes were found to shift the Page time
\cite{Yu:2025euq}. This suggests that the Page transition can be sensitive to quantum
corrections in the black hole dynamics. What is still missing is the corresponding real-time
description in a charged near-AdS$_2$ system. Soft modes are known to correct
near-extremal black hole thermodynamics and real-time fluxes, while charged islands are usually studied on a prescribed semiclassical background. We therefore ask how the soft mode quantum effects modify the evaporation law and how this modification shifts the Page curve.

In this paper we study charged near-AdS$_2$ black holes coupled to a non-gravitating bath
at fixed temperature and chemical potential. The bath can exchange both energy and charge
with the black hole\footnote{Bath constructions also give a useful way to make the radiation subsystem explicit
\cite{Rozali:2019day,Chen:2019uhq}, with related extensions in which the bath itself is
gravitating \cite{Anderson:2020vwi}.}. The black hole side is described by the charged near-AdS$_2$ boundary
effective theory, in which the Schwarzian mode and the $U(1)$ phase mode appear in the
same low-energy action \cite{Sachdev:2019bjn}. This gives a simple open system in which the evaporation dynamics and the Page curve can be treated in the same model.

We first derive the classical balance equations for the effective temperature $T(t)$ and
chemical potential $\mu(t)$. We then include the quantum fluctuations of the Schwarzian
mode and the $U(1)$ phase mode by replacing the boundary energy and the outgoing flux
with their averages over the soft modes. This gives quantum-corrected balance equations for the
real-time evaporating background. Compared with the classical equations, the corrected
equations contain new linear terms in the temperature. These terms modify the relaxation
toward the bath and also change the transient heating regime driven by a chemical potential
mismatch.

We then evaluate the no-island and island entropies on this quantum-corrected
evaporating background. The island formula itself is not modified. The correction enters
through the time-dependent map and the effective dilaton determined by the corrected
evaporation law. In this way the change of the Page curve is tied directly to the corrected
real-time dynamics. At first order, the Page time shift follows from the change in the
relative position of the two entropy branches near their crossing.

The main result is that the two soft modes affect the Page dynamics in different ways.
In the regime where our approximation is valid, the $1/K$ correction from the $U(1)$ phase
mode delays the Page transition. The Schwarzian contribution is more subtle: its sign is
controlled by an integral criterion, and in the numerical region where the near-AdS$_2$
low-temperature condition is satisfied, it tends to move the Page transition earlier. The
total Page time shift is therefore fixed by the competition between the Schwarzian sector
and the charge sector. This competition becomes visible in the scans over the $(C,K)$ plane.

The rest of this paper is organized as follows. In section \ref{sec:setup}, we review the
charged near-AdS$_2$ setup and the corresponding boundary effective theory. In
section \ref{section3}, we derive the classical balance equations for the effective temperature
and chemical potential, and then include soft mode quantum fluctuations to obtain their corrected equations. We also discuss the physical regimes of the corrected
evaporation equations. In section \ref{sec:page-curve}, we use the corrected background to formulate the no-island and island entropies. We then derive the leading
Page time shift and compare it with numerical Page curves and parameter scans. In section \ref{discussion}, we summarize the results and discuss possible extensions. Some technical details are presented in two appendices.

\section{Charged {$\mathrm{AdS}_2$} black holes and boundary effective theory}
\label{sec:setup}
We now introduce the charged AdS2 black hole and its boundary effective theory. We start from dilaton gravity coupled to a $U(1)$ gauge field and then review the corresponding boundary effective theory. In the near-AdS$_2$ regime, the low-energy dynamics is described by two boundary soft modes: the Schwarzian reparametrization mode $f$ and a $U(1)$ phase mode $\sigma$. We also couple the system to a non-gravitating bath, since later the bath will carry away both energy and charge.

\subsection{Setup}
We take a two-dimensional dilaton gravity theory \cite{Jackiw:1984je, Teitelboim:1983ux, Almheiri:2014cka} coupled to a $U(1)$ gauge field. The action contains the topological term, the bulk dilaton-Maxwell part, the Gibbons-Hawking term, the counterterm, and the matter sector:
\begin{equation}
S=
S_{\rm top}
+S_{\rm bulk}
+S_{\rm GH}
+S_{\rm ct}
+S_{\rm matter}\,,   
\end{equation}
with
\begin{align}
S_{\rm top}
&=
\frac{\Phi_0}{16\pi G_2}
\left[
\int_\mathcal{M} d^2x\,\sqrt{-g}\,R
+
2\int_{\partial \mathcal{M}} d\tau\,\sqrt{-\gamma}\,K
\right],  \\
S_{\rm bulk}
&=
\frac{1}{16\pi G_2}
\int_\mathcal{M} d^2x\,\sqrt{-g}\,
\left[
\Phi\!\left(R+\frac{2}{L^2}\right)
-\frac14\, Z(\Phi)\,F_{\mu\nu}F^{\mu\nu}
\right], \\
S_{\rm GH}
&=
\frac{1}{8\pi G_2}
\int_{\partial \mathcal{M}} d\tau\,\sqrt{-\gamma}\,\Phi K,
\qquad
S_{\rm ct}
=
-\frac{1}{8\pi G_2}
\int_{\partial \mathcal{M}} d\tau\,\sqrt{-\gamma}\,\frac{\Phi}{L}.
\end{align}

Here $S_{\rm top}$ gives the extremal entropy and does not affect the near-extremal
dynamics. The dynamical part is contained in $S_{\rm bulk}$, where the dilaton controls
the departure from extremality and the Maxwell term describes the charge sector. The
boundary terms $S_{\rm GH}$ and $S_{\rm ct}$ are included for a well-defined variational
principle and a finite on-shell action. We also include a matter sector $S_{\rm matter}$,
which will later provide the energy and charge fluxes.

In this paper we consider a charged AdS$_2$ black hole coupled to a non-gravitating bath \cite{DeVuyst:2022bua, Hollowood:2020cou, Chen:2020jvn}.  More precisely, we impose transparent boundary conditions for the matter field at the interface. For concreteness, the matter field is chosen as a free massless charged complex scalar,
\begin{equation}
S_{\rm matter}= - \int d^2x \sqrt{-g}\, g^{\mu\nu}(D_\mu\phi)^\dagger D_\nu\phi,
\qquad D_\mu=\nabla_\mu-iqA_\mu ,
\end{equation}
where $\phi$ has charge $q$.

\subsection{Boundary action: Schwarzian plus $U(1)$ phase modes}
Let us recall how the boundary action arises \cite{Almheiri:2014cka,Maldacena:2016upp,Jensen:2016pah,Engelsoy:2016xyb,Kitaev:2018wpr,Iliesiu:2019xuh}. The Schwarzian sector gives the universal low-energy dynamics of near-AdS$_2$ gravity, while the charged near-extremal theory contains an additional compact $U(1)$ boundary mode. This charged soft sector can be understood either from the near-extremal black hole thermodynamics or from the JT theory coupled to a two-dimensional gauge field \cite{Iliesiu:2020qvm,Kapec:2019ecr}. In the near-AdS$_2$ regime, the metric is locally AdS$_2$, while the Maxwell field has no local propagating degree of freedom. The remaining low-energy degrees of freedom are therefore boundary modes.

\paragraph{Schwarzian mode.}
Consider Euclidean AdS$_2$ in Poincar\'e coordinates,
\begin{equation}
ds^2=\frac{d\tau^2+dZ^2}{Z^2}\, .
\end{equation}
We denote a curve to regulate the asymptotic boundary
\begin{equation}
(\tau,Z)=(F(u),Z(u)) ,
\end{equation}
with $u$ the physical boundary time. We then set the standard asymptotically AdS boundary condition and fix the dilaton field asymptotics
\begin{equation}
\frac{F'(u)^2+Z'(u)^2}{Z(u)^2}
=
\frac{1}{\epsilon^2},
\qquad
\Phi\big|_{\partial M}
=
\Phi_b
=
\frac{\phi_r}{\epsilon},
\label{eq:ads2_bc}
\end{equation}
where $\epsilon\to0$ is the cutoff and $\phi_r$ is held fixed. Solving the first condition in small  $\epsilon$ gives
\begin{equation}
Z(u)=\epsilon F'(u)+\mathcal{O}(\epsilon^3),
\end{equation}
so the boundary curve is determined by a single reparametrization function $F(u)$. Evaluating the extrinsic curvature on this curve, one can obtain the following equation
\begin{equation}
K = 1+\epsilon^2 \{F(u),u\}+\mathcal{O}(\epsilon^4),
\label{eq:K_expand}
\end{equation}
where
\begin{equation}
\{F,u\}\equiv \frac{F'''(u)}{F'(u)}-\frac{3}{2}\left(\frac{F''(u)}{F'(u)}\right)^2
\end{equation}
is the Schwarzian derivative. Substituting \eqref{eq:ads2_bc} and \eqref{eq:K_expand} into the boundary terms $S_{\rm GH}+S_{\rm ct}$ yields the following Schwarzian action \cite{Maldacena:2016upp}
\begin{equation}
I_{\rm Sch}[F]
=
-C\int du\,\{F(u),u\},
\qquad
C\equiv\frac{\phi_r}{8\pi G_2}.
\end{equation}
Thus the gravitational boundary dynamics is encoded by a single reparametrization $F$. See the relevant review \cite{Mertens:2022irh} for more details. For the black hole studied here, it is convenient to write the Schwarzian action in the standard thermal form with a circle of length $\beta$,
\begin{equation}
I_{\rm Sch}[F]
=-C\int d\tau\,\{F(\tau),\tau\}\equiv
-C\int d\tau\,
\left\{
\tan\!\left(\frac{\pi}{\beta}f(\tau)\right),\tau
\right\},
\end{equation}
with the following periodicity
\begin{equation}
f(\tau+\beta)=f(\tau)+\beta .
\end{equation}
\paragraph{$U(1)$ phase mode.}
The gauge sector behaves similarly. In two dimensions, the Maxwell equation
\begin{equation}
\nabla_\mu\!\bigl(Z(\Phi)F^{\mu\nu}\bigr)=0
\end{equation}
does not describe a local propagating degree of freedom. The only physical information are a conserved electric flux and a global Wilson-line degree of freedom along the Euclidean boundary circle, i.e.,  a boundary holonomy. In the grand-canonical ensemble,  this holonomy is conveniently described by a compact boundary phase field $\sigma(\tau)$ \cite{Sachdev:2019bjn, Iliesiu:2019lfc}, defined up to a constant
shift. Gauge invariance then implies that the phase mode can only enter through the combination
$\sigma'(\tau)-i\mu f'(\tau)$, and its effective action is given by 
\begin{equation}
I_{U(1)}[\sigma;f]
=
\frac{K}{2}\int d\tau \,
\bigl(\sigma'(\tau)-i\mu f'(\tau)\bigr)^2 .
\end{equation}
Here $K$ is the charge susceptibility (or compressibility) of the near-extremal black hole. The
phase mode is compact,
\begin{equation}
\sigma \sim \sigma+2\pi ,
\end{equation}
and therefore on the Euclidean thermal circle one must sum over winding sectors,
\begin{equation}
\sigma(\tau+\beta)=\sigma(\tau)+2\pi m,
\qquad m\in\mathbb Z .
\end{equation}
In order to decouple Schwarzian mode and $U(1)$ phase mode, it is often convenient to remove the explicit $\mu$-dependence by the field redefinition
\begin{equation}
\tilde{\sigma}(\tau)\equiv \sigma(\tau)-i\mu f(\tau)\,.
\label{redef}
\end{equation}
The corresponding twisted thermal boundary condition becomes
\begin{equation}
\tilde{\sigma}(\tau+\beta)
=
\tilde{\sigma}(\tau)+2\pi m-i\mu\beta,
\qquad
m\in\mathbb Z.
\end{equation}
In terms of $\tilde{\sigma}$ the Euclidean boundary action becomes
\begin{equation}
I_{\rm bdy}[f,\tilde{\sigma}]
=
-C\int d\tau\,
\left\{
\tan\!\left(\frac{\pi}{\beta}f(\tau)\right),\tau
\right\}
+
\frac{K}{2}\int d\tau\,\tilde{\sigma}'(\tau)^2 .
\end{equation}
In the following we will drop the tilde and still denote the phase mode by $\sigma$. In the Lorentzian signature, the low-energy boundary theory is therefore
\begin{equation}
I_{\rm bdy}[F,\sigma]
=
-C\int d t\,
\left\{
F(t), t
\right\}
+
\frac{K}{2}\int d t\,\sigma'(t)^2 ,
\end{equation}
This gives the boundary effective theory used in the following sections.

\subsection{Static charged black hole solution and thermodynamic dictionary}
\label{scbhsatd}
For later use, we also describe the static charged saddle and the corresponding thermodynamic dictionary. In Euclidean signature, after the field redefinition in \eqref{redef}, the action factorizes, and the static saddle is simply
\begin{equation}
f(\tau)=\tau,
\qquad
\tilde{\sigma}'(\tau)=0 .
\end{equation}
Equivalently, in the original variables this corresponds to a constant chemical potential
background,
\begin{equation}
\sigma'(\tau)=i\mu f'(\tau)=i\mu .
\end{equation}
After analytic continuation to Lorentzian time, the corresponding static charged black hole
is described by
\begin{equation}
F(t)=\frac{\beta}{\pi}\tanh\!\left(\frac{\pi}{\beta}t\right),
\qquad
\dot{\sigma}(t)=\mu ,
\label{eq:charged_static_lorentzian_boundary}
\end{equation}
where $T=\beta^{-1}$ is the Hawking temperature and $\mu$ is the chemical potential.

The boundary reparametrization \eqref{eq:charged_static_lorentzian_boundary} lifts to the
two-dimensional bulk in the standard way by introducing null coordinates
\begin{equation}
X^+=F(u),
\qquad
X^-=F(v).
\end{equation}
The metric is the universal near-AdS$_2$ metric associated with the frame $F$,
\begin{equation}
ds^2
=
-\frac{4\,F'(u)F'(v)}{\bigl(F(u)-F(v)\bigr)^2}\,du\,dv .
\end{equation}
For the static saddle \eqref{eq:charged_static_lorentzian_boundary}, this becomes
\begin{equation}
ds^2
=
-\frac{4(\pi T)^2}{\sinh^2\!\bigl[\pi T (u-v)\bigr]}\,du\,dv ,
\label{eq:charged_static_metric}
\end{equation}
which is the standard AdS$_2$ black hole metric at Hawking temperature $T$. The associated
dilaton profile is
\begin{equation}
\Phi(u,v)
=
2\phi_r\pi T\coth\!\bigl[\pi T (u-v)\bigr],
\label{eq:charged_static_dilaton_uv}
\end{equation}
or, equivalently, in Poincar\'e coordinates,
\begin{equation}
\Phi(X^+,X^-)
=
2\phi_r\,
\frac{1-(\pi T)^2 X^+X^-}{X^+-X^-}.
\end{equation}
The future and past horizons are located at
\begin{equation}
u=+\infty,
\qquad
v=-\infty,
\end{equation}
or equivalently at
\begin{equation}
X^\pm=\pm \frac{1}{\pi T}.
\end{equation}

The metric in \eqref{eq:charged_static_metric} and the dilaton profile in \eqref{eq:charged_static_dilaton_uv} have the same form as
in neutral JT gravity. This is because, in the low-temperature near-AdS$_2$ regime, the
geometry remains locally AdS$_2$, while the charge sector mainly changes the relation
between $(T,\mu)$ and $(E,Q)$. Away from this regime, the full charged solution does
depend on the detailed form of $Z(\Phi)$.

To extract this thermodynamic dictionary, it is convenient to pass from a single classical saddle to the grand-canonical partition function of the same low-energy theory. This use of the exact or semiclassical low-energy density of states follows the standard JT gravity description of near-AdS$_2$ quantum gravity~\cite{Saad:2019lba,Kitaev:2018wpr,Iliesiu:2019xuh}, now supplemented by the $U(1)$ charge sectors of the boundary gauge theory~\cite{Iliesiu:2020qvm,Kapec:2019ecr}. Since the JT model coupled to 2d Maxwell theory can also be described by the 2d $U(1)$ BF model, the charged thermal partition function takes the following form
\begin{equation}
Z(\beta,\mu)
=
\sum_{Q\in e\mathbb{Z}}
\int_{Q^2/2K}^{\infty}
dE\,
e^{-\beta(E-\mu Q)}
\sinh\!\left(
2\pi\sqrt{2C}\sqrt{E-\frac{Q^2}{2K}}
\right)\,,
\end{equation}
up to an overall $(\beta,\mu)$ independent normalization constant. In the semiclassical regime, one may approximate $\sinh x \sim \tfrac12 e^x$ and write
\begin{equation}
Z(\beta,\mu)\sim
\sum_Q \int dE\,
\exp\!\Bigl[
S(E,Q)-\beta(E-\mu Q)
\Bigr],
\qquad
S(E,Q)=2\pi\sqrt{2C}\sqrt{E-\frac{Q^2}{2K}} .
\end{equation}
The saddle point equations are therefore
\begin{equation}
\frac{\partial S}{\partial E}=\beta,
\qquad
\frac{\partial S}{\partial Q}=-\beta\mu ,
\end{equation}
which give
\begin{equation}
\beta
=
\frac{\pi\sqrt{2C}}{\sqrt{E-Q/2K}},
\qquad
\mu=\frac{Q}{K}.
\end{equation}
It is therefore natural to identify the thermodynamic variables through
\begin{equation}
E(T,\mu)=2\pi^2 C T^2+\frac{K\mu^2}{2},
\qquad
Q(T,\mu)=K\mu .
\end{equation}
Equivalently, one can express $E$ in terms of $Q$ as
\begin{equation}
E(T,Q)=2\pi^2 C T^2+\frac{Q^2}{2K},
\qquad
T=\frac{1}{\pi\sqrt{2C}}\sqrt{E-\frac{Q^2}{2K}} .
\end{equation}

The Bekenstein--Hawking entropy then follows from the first law,
\begin{equation}
dE=TdS_{\rm BH}+\mu dQ ,
\end{equation}
which yields
\begin{equation}
S_{\rm BH}(T,\mu)=4\pi^2 C\,T ,
\label{eq:charged_SBH_Tmu}
\end{equation}
or, equivalently,
\begin{equation}
S_{\rm BH}(E,Q)
=
2\pi\sqrt{2C}\sqrt{E-\frac{Q^2}{2K}} .
\end{equation}
In particular, the extremal limit is given by
\begin{equation}
E=\frac{Q^2}{2K},
\qquad
T=0 .
\end{equation}

Two conditions should be distinguished here. The first is the low-temperature, or near-extremal regime,
\begin{equation}
TL \ll 1 .
\end{equation}
The second is the validity of the semiclassical saddle, which requires
\begin{equation}
S_{\rm BH}\gg 1 .
\end{equation}
Using \eqref{eq:charged_SBH_Tmu}, this becomes
\begin{equation}
4\pi^2 C T \gg 1 .
\end{equation}
Hence, we will work in the overlap region
\begin{equation}
TL\ll 1,\qquad 4\pi^2 C T \gg 1 .
\label{semiclassical_low}
\end{equation}
This completes the thermodynamic dictionary of the black hole side. We next turn to the real-time evaporation process.

\section{Evaporation dynamics of the charged black hole}
\label{section3}
\subsection{Matter fluxes and classical balance equations}
\label{matterfluxes}
Let us now study the real time evaporation.  The black hole exchanges energy and charge with the bath through transparent boundary conditions at the interface $u=v=t$. To obtain the balance equations, it is not necessary to know the full bulk solution. We only need the renormalized chiral one-point functions of the charged matter fields in the background fixed by the boundary modes $(F,\sigma)$.

For the charged scalar introduced in section \ref{sec:setup}, the reparametrization mode determines the
chiral coordinates through
\begin{equation}
X^+ = F(u),\qquad X^- = F(v)\,,
\end{equation}
while boundary phase mode gives the gauge dressing of charged operators. After the normal ordering defined with respect to the Poincar\'e vacuum, the renormalized outgoing one-point functions are\footnote{These expressions follow from the standard conformal transformation of the chiral stress tensor and the gauge dressing of charged matter
operators. A more explicit point-splitting derivation can be found in section 5 of \cite{DeVuyst:2022bua}.}
\begin{equation}
\langle :T_{uu}(t): \rangle_{\rm out}
= -\frac{c}{24\pi}\{F(t),t\}
+ \frac{q^2}{4\pi}\dot{\sigma}(t)^2,
\qquad
\langle :J_u(t): \rangle_{\rm out}
= \frac{q^2}{2\pi}\dot{\sigma}(t).
\label{eq:classical-outgoing-flux}
\end{equation}
The same expressions hold in the $v$ sector. In the net flux through the interface, the
frame-dependent and gauge-dependent inhomogeneous terms cancel between the two chiral
components. The balance equation can therefore be written directly in terms of the
normal-ordered fluxes,
\begin{equation}
\langle T_{vv}(t)\rangle-\langle T_{uu}(t)\rangle
=
\langle :T_{vv}(t):\rangle-\langle :T_{uu}(t):\rangle,
\end{equation}
and similarly for the charge current.

The black hole energy and charge in the classical boundary theory are
\begin{equation}
    E_{\rm BH}(t)
    =
    -C\{F(t),t\}
    +
    \frac{K}{2}\dot{\sigma}(t)^2,
    \qquad
    Q_{\rm BH}(t)
    =
    K\dot{\sigma}(t).
\end{equation}
For the static saddle
\begin{equation}
    F(t)=\frac{\beta}{\pi}\tanh\frac{\pi t}{\beta},
    \qquad
    \dot{\sigma}(t)=\mu,
\end{equation}
this reproduces the thermodynamic dictionary
\begin{equation}
    E_{\rm BH}
    =
    2\pi^2 C T^2+\frac{K}{2}\mu^2,
    \qquad
    Q_{\rm BH}=K\mu,
    \qquad
    T=\beta^{-1}.
\end{equation}

The bath is kept in a stationary grand-canonical state with temperature
$T_b=\beta_b^{-1}$ and chemical potential $\mu_b$. Its incoming one-point
functions at the interface are
\begin{equation}
    \langle :T_{vv}(t): \rangle_{\rm in}
    =
    \frac{\pi c}{12}T_b^2
    +
    \frac{q^2}{4\pi}\mu_b^2,
    \qquad
    \langle :J_v(t): \rangle_{\rm in}
    =
    \frac{q^2}{2\pi}\mu_b .
    \label{eq:classical-incoming-flux}
\end{equation}
Transparent boundary conditions then give the classical balance laws
\begin{equation}
    \frac{dE_{\rm BH}}{dt}
    =
    \langle :T_{vv}(t): \rangle_{\rm in}
    -
    \langle :T_{uu}(t): \rangle_{\rm out},
    \qquad
    \frac{dQ_{\rm BH}}{dt}
    =
    \langle :J_v(t): \rangle_{\rm in}
    -
    \langle :J_u(t): \rangle_{\rm out}.
\end{equation}
Using \eqref{eq:classical-outgoing-flux} and
\eqref{eq:classical-incoming-flux}, we obtain the classical balance equations with fixed $(F,\sigma)$ background:
\begin{equation}
    \frac{d}{dt}
    \left[
    -C\{F(t),t\}
    +
    \frac{K}{2}\dot{\sigma}(t)^2
    \right]
    =
    \frac{\pi c}{12}T_b^2
    +
    \frac{q^2}{4\pi}\mu_b^2
    +
    \frac{c}{24\pi}\{F(t),t\}
    -
    \frac{q^2}{4\pi}\dot{\sigma}(t)^2,
\end{equation}
and
\begin{equation}
    K\ddot{\sigma}(t)
    =
    -\frac{q^2}{2\pi}
    \left(\dot{\sigma}(t)-\mu_b\right).
\end{equation}

It is convenient to rewrite these equations in terms of the effective temperature
and chemical potential of the black hole,
\begin{equation}
    \{F(t),t\}
    =
    -2\pi^2 T(t)^2,
    \qquad
    \mu(t)\equiv \dot{\sigma}(t).
\end{equation}
The balance laws then become
\begin{equation}
    \frac{d}{dt}
    \left(
    2\pi^2 C T(t)^2
    +
    \frac{K}{2}\mu(t)^2
    \right)
    =
    -\frac{\pi c}{12}
    \left(T(t)^2-T_b^2\right)
    -
    \frac{q^2}{4\pi}
    \left(\mu(t)^2-\mu_b^2\right),
    \label{eq:classical-energy-balance-Tmu}
\end{equation}
and
\begin{equation}
    K\dot{\mu}(t)
    =
    -\frac{q^2}{2\pi}
    \left(\mu(t)-\mu_b\right).
    \label{eq:classical-charge-balance-Tmu}
\end{equation}
These coupled first order equations can be solved explicitly. Let us denote the initial data when the black hole is put in contact with the bath
\begin{equation}
    T(0)=T_0,
    \qquad
    \mu(0)=\mu_0.
\end{equation}
The charge equation is solved by
\begin{equation}
    \mu(t)
    =
    \mu_b+(\mu_0-\mu_b)e^{-\gamma_Q t},
    \qquad
    \gamma_Q\equiv \frac{q^2}{2\pi K}.
    \label{eq:classical-mu-solution}
\end{equation}
Equivalently,
\begin{equation}
    Q_{\rm BH}(t)
    =
    K\mu_b+
    \bigl(Q_{\rm BH}(0)-K\mu_b\bigr)e^{-\gamma_Q t}.
\end{equation}
We interpret $\gamma_Q$ as the relaxation rate of the charge sector. It is convenient to eliminate $\dot{\mu}(t)$ and rewrite the energy equation as a closed equation for $T(t)^2$. Taking the time derivative in \eqref{eq:classical-energy-balance-Tmu} gives
\begin{equation}
2\pi^2 C\,\frac{d}{dt}T(t)^2+K\mu(t)\dot{\mu}(t)
=
-\frac{\pi c}{12}\bigl(T(t)^2-T_b^2\bigr)
-\frac{q^2}{4\pi}\bigl(\mu(t)^2-\mu_b^2\bigr).
\end{equation}
Using \eqref{eq:classical-charge-balance-Tmu} to eliminate $K\dot{\mu}(t)$, one finds a closed equation for $T(t)^2$,
\begin{equation}
2\pi^2 C\,\frac{d}{dt}T(t)^2
=
-\frac{\pi c}{12}\bigl(T(t)^2-T_b^2\bigr)
+\frac{q^2}{4\pi}\bigl(\mu(t)-\mu_b\bigr)^2 .
\label{eq:T2_closed_classical}
\end{equation}
Substituting \eqref{eq:classical-mu-solution} into \eqref{eq:T2_closed_classical}, and defining
\begin{equation}
\delta T^2(t)\equiv T(t)^2-T_b^2\,,
\qquad
\delta\mu_0\equiv \mu_0-\mu_b,
\end{equation}
we obtain
\begin{equation}
\frac{d}{dt}\delta T^2(t)+k\,\delta T^2(t)
=
\frac{q^2}{8\pi^3 C}\,\delta\mu_0^{\,2}\,e^{-2\gamma_Q t}\,,\qquad k \equiv \frac{c}{24\pi C}\,.
\end{equation}
For the generic case $k\neq 2\gamma_Q$, the solution is
\begin{equation}
T(t)^2
=
T_b^2
+
\left[
T_0^2-T_b^2
-\frac{q^2\,\delta\mu_0^{\,2}}{8\pi^3 C\,(k-2\gamma_Q)}
\right]e^{-kt}
+
\frac{q^2\,\delta\mu_0^{\,2}}{8\pi^3 C\,(k-2\gamma_Q)}\,e^{-2\gamma_Q t}.
\end{equation}
Here $k$ is the evaporation rate in the neutral case. For the resonant case $k=2\gamma_Q$, both exponentials have the same decay rate and the solution becomes
\begin{equation}
T(t)^2
=
T_b^2
+
\bigl(T_0^2-T_b^2\bigr)e^{-kt}
+
\frac{q^2\,\delta\mu_0^{\,2}}{8\pi^3 C}\, t\,e^{-kt}.
\end{equation}
In the following, we discuss several physical limits and regimes.
\begin{itemize}
\item $\mu_0=\mu_b$
    
In this physical limit, the chemical potential mismatch vanishes, so the $U(1)$ source term in \eqref{eq:T2_closed_classical} drops out and one recovers the neutral result
\begin{equation}
T(t)^2-T_b^2
=
\bigl(T_0^2-T_b^2\bigr)e^{-kt}.
\end{equation}
Thus, at the classical level, the charge sector affects the energy evolution only through the chemical potential mismatch $\mu(t)-\mu_b$.
\item $\mu_0-\mu_b\neq 0$

For nonzero chemical potential mismatch, the charge sector can change the temperature evolution. This is obvious in the right hand side of classical equation \eqref{eq:T2_closed_classical}. To see this more clearly, let
\begin{equation}
Y(t)\equiv T(t)^2,\qquad
B\equiv T_0^2-T_b^2\,,
\end{equation}
and define
\begin{equation}
A\equiv \frac{q^2(\mu_0-\mu_b)^2}{8\pi^3 C},
\qquad
k\equiv \frac{c}{24\pi C}.
\end{equation}
Then the classical temperature equation can be written as
\begin{equation}
\dot Y+k(Y-T_b^2)=A e^{-2\gamma_Q t}.
\end{equation}
Here we already substitute the charge solution \eqref{eq:classical-mu-solution}. The positive source term on the right hand side can temporarily increase the near-extremal thermal energy $2\pi^2 C T^2$, even while the total black hole energy is decreasing.

In particular, under our black hole evaporation setup $T_0>T_b$, the temperature can initially increases rather than decreases if
\begin{equation}
A>kB,
\end{equation}
which means
\begin{equation}
\frac{q^2}{4\pi}(\mu_0-\mu_b)^2>\frac{\pi c}{12}(T_0^2-T_b^2).
\end{equation}
When this condition is satisfied, there is a unique turning time $t_M$ at which
$\dot T(t_M)=0$. For $k\neq 2\gamma_Q$, it is given by
\begin{equation}
t_M=
\frac{1}{k-2\gamma_Q}
\log\left[
\frac{k\bigl(A-(k-2\gamma_Q)B\bigr)}{2\gamma_Q A}
\right].
\end{equation}
In the resonant case $k=2\gamma_Q$, the corresponding expression is
\begin{equation}
t_M=\frac{A-kB}{kA}.
\end{equation}
Thus, for $0<t<t_M$, the temperature monotonically increases , while for $t>t_M$ the temperature monotonically decreases toward the bath temperature $T_b$.

\item $\gamma_Q \gg k$

In this parameter regime, the chemical potential equilibrates rapidly. Once this fast charge relaxation stage is over, the temperature approximately follows the neutral relaxation law:
\begin{equation}
T(t)^2 \simeq T_b^2 + \bigl(T_0^2-T_b^2\bigr)e^{-kt},
\qquad
t\gg \gamma_Q^{-1}.
\end{equation}

\item $T_b=0\,,\mu_b=0$

For this purely absorbing bath, the classical solution reduces to
\begin{equation}
\mu(t)=\mu_0 e^{-\gamma_Q t},
\end{equation}
and for $k\neq 2\gamma_Q$
\begin{equation}
T(t)^2
=
\left[
T_0^2-\frac{q^2\mu_0^2}{8\pi^3 C\,(k-2\gamma_Q)}
\right]e^{-kt}
+
\frac{q^2\mu_0^2}{8\pi^3 C\,(k-2\gamma_Q)}\,e^{-2\gamma_Q t}\,.
\end{equation}
It should be emphasized that the late-time limit of the classical solution must be treated with care. The thermodynamic dictionary used in this paper relies on the semiclassical near-extremal black hole saddle, whose regime of validity requires
\begin{equation}
4\pi^2 C\,T(t)\gg 1,
\qquad
T(t)\ll 1.
\end{equation}
Therefore, although the purely absorbing bath solution formally drives $T(t)$ towards zero, the effective description we used here should only be trusted while the evolution remains inside this overlap regime. Once
\begin{equation}
T(t)\sim \mathcal{O}(C^{-1}),
\end{equation}
the black hole entropy is no longer parametrically large, and the present semiclassical calculation is no longer quantitatively controlled. For this reason, the solution should be understood as an effective description of the near-extremal region not a controlled way to exact extremal region. 
\end{itemize}
\subsection{Quantum-corrected balance equations}
\label{qcbe}
The next step is to include the quantum fluctuations of the two boundary soft modes. We work in the adiabatic regime. At each time the black hole is described by an effective temperature and an effective chemical potential. The classical balance equations \eqref{eq:classical-energy-balance-Tmu} \eqref{eq:classical-charge-balance-Tmu} are promoted to expectation-value equations by replacing the classical boundary energy and fluxes with their averages over the soft modes. We take the resulting expectation-value equations as the quantum-corrected balance equations. The corresponding quantum-corrected energy is obtained from the exact soft-mode partition function,
\begin{equation}
E_{\rm BH}^{\rm q}(\beta,\mu)
=
-\partial_\beta \log Z_{\rm soft}(\beta,\mu)
+
\mu\,\frac{1}{\beta}\partial_\mu \log Z_{\rm soft}(\beta,\mu)\,.
\end{equation}
Since the Schwarzian mode and the $U(1)$ phase mode are decoupled from each other, we can write
\begin{equation}
Z_{\rm soft}(\beta,\mu)=Z_{\rm Sch}(\beta)\,Z_{U(1)}(\beta,\mu)\,,
\end{equation}
and
\begin{equation}
E_{\rm BH}^{\rm q}(\beta,\mu)=E_{\rm Sch}(\beta)+E_{U(1)}(\beta,\mu)\,.
\end{equation}
For the Schwarzian mode, the one-loop exact partition function \cite{Stanford:2017thb, Mertens:2022irh} is given by
\begin{equation}
Z_{\rm Sch}(\beta)
=
\frac{1}{4\pi^2}
\left(\frac{2\pi C}{\beta}\right)^{3/2}
\exp\!\left(\frac{2\pi^2 C}{\beta}\right),
\end{equation}
Using
\begin{equation}
E_{\rm Sch}(\beta)=-\partial_\beta \log Z_{\rm Sch}(\beta),
\end{equation}
one obtains
\begin{equation}
E_{\rm Sch}(\beta)
=
\frac{2\pi^2 C}{\beta^2}+\frac{3}{2\beta}
=
2\pi^2 C T^2+\frac{3}{2}T.
\label{eq:ESch}
\end{equation}
For the $U(1)$ phase mode, the partition function \cite{Sachdev:2019bjn, Iliesiu:2019lfc} is given by
\begin{equation}
Z_{U(1)}(\beta,\mu)
=
\sum_{n\in\mathbb Z}
\exp\!\left[-\beta\left(\frac{n^2}{2K}-\mu n\right)\right].
\end{equation}
In the regime $\beta \ll K$, or equivalently $T\gg K^{-1}$, Poisson resummation gives
\begin{equation}
\log Z_{U(1)}(\beta,\mu)
=
\frac{\beta K\mu^2}{2}
+
\frac12 \log\!\left(\frac{2\pi K}{\beta}\right)
+
O\!\left(e^{-2\pi^2K/\beta}\right).
\end{equation}
From the grand canonical thermodynamic relations
\begin{equation}
Q_{U(1)}=\frac{1}{\beta}\partial_\mu \log Z_{U(1)}=K\mu,
\end{equation}
and
\begin{equation}
E_{U(1)}(\beta,\mu)
=
-\partial_\beta \log Z_{U(1)}+\mu Q_{U(1)},
\end{equation}
we obtain
\begin{equation}
E_{U(1)}(\beta,\mu)
=
\frac{K}{2}\mu^2+\frac{1}{2\beta}
=
\frac{K}{2}\mu^2+\frac12 T.
\label{eq:EU1}
\end{equation}
Combining \eqref{eq:ESch} and \eqref{eq:EU1}, we obtain
\begin{equation}
E_{\rm BH}^{\rm q}(T,\mu)
=
2\pi^2 C T^2+\frac32 T+\frac{K}{2}\left(\mu^2+\frac{T}{K}\right)
=
2\pi^2 C T^2+2T+\frac{K}{2}\mu^2.
\label{eq:quantum_energy_soft}
\end{equation}
Compared with the classical result, the quantum-corrected energy contains an additional
term linear in $T$. The Schwarzian sector contributes $\frac{3}{2}T$, while the $U(1)$
sector contributes $\frac{1}{2}T$.

We now turn to the fluxes. In subsection \ref{matterfluxes}, the outgoing one-point functions were evaluated in a fixed background $(F,\sigma)$. The quantum-corrected fluxes are obtained by taking the soft-mode average of the same observables. In particular, the outgoing energy flux is
\begin{equation}
\langle {:}T_{uu}(t){:}\rangle_{\rm out}^{\rm q}
=
-\frac{c}{24\pi}\,\langle \{F(t),t\}\rangle
+
\frac{q^2}{4\pi}\,\langle \dot{\sigma}(t)^2\rangle .
\end{equation}
We now evaluate the two expectation values separately.

For the Schwarzian part, the exact thermal energy of the reparametrization mode is
\begin{equation}
E_{\rm Sch}(T)=-C\,\langle \{F(t),t\}\rangle
=
2\pi^2 C T^2+\frac{3}{2}T.
\end{equation}
Therefore,
\begin{equation}
\langle \{F(t),t\}\rangle
=
-2\pi^2 T(t)^2-\frac{3}{2C}T(t),
\end{equation}
and hence
\begin{equation}
-\frac{c}{24\pi}\,\langle \{F(t),t\}\rangle
=
\frac{\pi c}{12}T(t)^2+\frac{c}{16\pi C}T(t).
\label{eq:Sch_flux_piece}
\end{equation}

For the $U(1)$ sector, the outgoing flux depends on $\dot{\sigma}^2$. In the quantum theory,
the phase mode is treated as a thermal rotor. Writing the charge as
\begin{equation}
Q=K\dot{\sigma},
\end{equation}
one has
\begin{equation}
\langle \dot{\sigma}(t)\rangle = \mu(t),
\qquad
\langle Q\rangle = K\mu(t).
\end{equation}
Moreover, in the grand-canonical ensemble the charge fluctuations are characterized by
\begin{equation}
\mathrm{Var}(Q)=KT,
\end{equation}
so that
\begin{equation}
\langle Q^2\rangle = \langle Q\rangle^2+\mathrm{Var}(Q)
=K^2\mu(t)^2+KT(t).
\end{equation}
Dividing by $K^2$, we obtain
\begin{equation}
\langle \dot{\sigma}(t)^2\rangle
=
\mu(t)^2+\frac{T(t)}{K}.
\end{equation}
Therefore the $U(1)$ contribution to the outgoing flux is
\begin{equation}
\frac{q^2}{4\pi}\,\langle \dot{\sigma}(t)^2\rangle
=
\frac{q^2}{4\pi}\left(\mu(t)^2+\frac{T(t)}{K}\right).
\label{eq:U1_flux_piece}
\end{equation}

Combining \eqref{eq:Sch_flux_piece} and \eqref{eq:U1_flux_piece}, we finally arrive at
\begin{equation}
\langle {:}T_{uu}(t){:}\rangle_{\rm out}^{\rm q}
=
\frac{\pi c}{12}T(t)^2
+
\frac{c}{16\pi C}T(t)
+
\frac{q^2}{4\pi}\left(\mu(t)^2+\frac{T(t)}{K}\right).
\label{eq:quantum_Tuu}
\end{equation}
This is the quantum-corrected version of the classical outgoing flux. Since the current is linear in $\dot{\sigma}$ already at the classical level, its quantum-corrected form is obtained simply by replacing $\dot{\sigma}(t)$ with its expectation value $\mu(t)$, namely
\begin{equation}
\langle {:}J_u(t){:}\rangle_{\rm out}^{\rm q}
=
\frac{q^2}{2\pi}\mu(t).
\end{equation}
The first term in \eqref{eq:quantum_Tuu} is the classical outgoing energy flux, the second is the Schwarzian correction, and the last term comes from the $U(1)$ sector after replacing the classical quantity $\dot{\sigma}^2$ by its quantum average,
\begin{equation}
\langle \dot{\sigma}^2 \rangle
=
\mu^2+\frac{T}{K}.
\end{equation}

For the bath, we keep the same grand-canonical state with fixed $(T_b,\mu_b)$. Therefore the incoming fluxes are
\begin{equation}
\langle {:}T_{vv}(t){:}\rangle_{\rm in}^{\rm q}
=
\frac{\pi c}{12}T_b^2
+
\frac{c}{16\pi C}T_b
+
\frac{q^2}{4\pi}\left(\mu_b^2+\frac{T_b}{K}\right),
\end{equation}
and
\begin{equation}
\langle {:}J_v(t){:}\rangle_{\rm in}^{\rm q}
=
\frac{q^2}{2\pi}\mu_b.
\label{eq:quantum_Jv}
\end{equation}

We now impose the same balance laws as before, but with the quantum-corrected quantities:
\begin{equation}
\frac{d}{dt}E_{\rm BH}^{\rm q}(t)
=
\langle {:}T_{vv}(t){:}\rangle_{\rm in}^{\rm q}
-
\langle {:}T_{uu}(t){:}\rangle_{\rm out}^{\rm q},
\end{equation}
\begin{equation}
\frac{d}{dt}Q_{\rm BH}^{\rm q}(t)
=
\langle {:}J_v(t){:}\rangle_{\rm in}^{\rm q}
-
\langle {:}J_u(t){:}\rangle_{\rm out}^{\rm q}.
\end{equation}
Substituting \eqref{eq:quantum_energy_soft}--\eqref{eq:quantum_Jv}, we obtain
\begin{equation}
\begin{aligned}
\frac{d}{dt}
\left(
2\pi^2 C T(t)^2+2T(t)+\frac{K}{2}\mu(t)^2
\right)
&=
-\frac{\pi c}{12}\bigl(T(t)^2-T_b^2\bigr)
-\frac{c}{16\pi C}\bigl(T(t)-T_b\bigr)
\\
&\quad
-\frac{q^2}{4\pi}
\left[
\mu(t)^2-\mu_b^2+\frac{T(t)-T_b}{K}
\right] .
\end{aligned}
\label{eq:quantum_balance_full}
\end{equation}
together with
\begin{equation}
K\dot{\mu}(t)
=
-\frac{q^2}{2\pi}\bigl(\mu(t)-\mu_b\bigr).
\label{eq:quantum_charge_eq}
\end{equation}

Equation \eqref{eq:quantum_balance_full} is the quantum-corrected energy balance equation that replaces the equation \eqref{eq:classical-energy-balance-Tmu} in the last subsection. The structure here is obvious: the balance law is still valid in form, but the black hole energy and the energy flux are now replaced by their quantum soft mode averages. 

We also rewrite \eqref{eq:quantum_balance_full} as an equation for $T(t)$ alone. Expanding the time derivative gives
\begin{equation}
\begin{aligned}
\bigl(4\pi^2 C T(t)+2\bigr)\dot{T}(t)
+K\mu(t)\dot{\mu}(t)
&=
-\frac{\pi c}{12}\bigl(T(t)^2-T_b^2\bigr)
-\frac{c}{16\pi C}\bigl(T(t)-T_b\bigr)
\\
&\quad
-\frac{q^2}{4\pi}
\left[
\mu(t)^2-\mu_b^2+\frac{T(t)-T_b}{K}
\right] .
\end{aligned}
\end{equation}
Using \eqref{eq:quantum_charge_eq} to eliminate $\dot{\mu}(t)$, one finds
\begin{equation}
\bigl(4\pi^2 C T(t)+2\bigr)\dot{T}(t)
=
-\frac{\pi c}{12}\bigl(T(t)^2-T_b^2\bigr)
-\frac{c}{16\pi C}\bigl(T(t)-T_b\bigr)
+\frac{q^2}{4\pi}\bigl(\mu(t)-\mu_b\bigr)^2
-\frac{q^2}{4\pi K}\bigl(T(t)-T_b\bigr).
\end{equation}

The matter sector is still the same quantum matter sector as before. However, the boundary soft modes are no longer treated as a fixed classical background. Once these modes are quantized, they modify both the black hole energy and the outgoing flux, leading to the $1/C$ and $1/K$ corrections that appear in the balance equation.

\subsection{Several physical limits and regimes of the quantum-corrected balance equations}
\label{subsec:qbalance-limits}
Now we discuss the physical meaning of the corrected equations. Compared with the classical evolution, there are two changes. First, the black hole energy has an additional linear temperature term,
\begin{equation}
E_{\rm BH}^{q}(T,\mu)=2\pi^{2}CT^{2}+2T+\frac{K}{2}\mu^{2},
\end{equation}
Second, the outgoing energy flux also has new linear temperature terms,
\begin{equation}
\frac{c}{16\pi C}(T-T_b),
\qquad
\frac{q^{2}}{4\pi K}(T-T_b),
\end{equation}
For the late time analysis below, the fixed point is assumed to lie in the controlled window
\begin{equation}
T_b L \ll 1,\qquad 4\pi^{2} C T_b \gg 1,\qquad K T_b \gg 1\,.
\end{equation}
Here the first two inequalities are the overlap region introduced in subsection \ref{scbhsatd}, while the last one is the regime in which the grand canonical $U(1)$ rotor admits the high temperature expansion used in subsection \ref{qcbe}. 

The corrected equations cannot be solved analytically in a closed form for general parameters. Nevertheless, two useful descriptions are available. First, the asymptotic late time behavior to the bath fixed point can be obtained directly by linearizing the balance equations.  Second, for the later Page time analysis, we need a controlled approximation.  In our paper's controlled window, this can be obtained by expanding the temperature profile around the classical solution in powers of $1/(CT)$ and $1/(KT)$.  The details of deriving this series solution are given in appendix \ref{appA} and we only use its final form in the following. We first consider the late-time near-equilibrium regime. To study the asymptotic approach to equilibrium, we write
\begin{equation}
T(t)=T_b+\delta T(t),\qquad \mu(t)=\mu_b+\delta\mu(t).
\end{equation}
The charge equation gives
\begin{equation}
\dot{\delta\mu}=-\frac{1}{\tau_Q}\delta\mu,
\qquad
\tau_Q\equiv \frac{2\pi K}{q^2},
\label{eq:deltamu-linear-33}
\end{equation}
while the energy equation takes the form
\begin{equation}
(4\pi^{2}CT_b+2)\dot{\delta T}+K\mu_b\dot{\delta\mu}
=
-\left(\frac{\pi c}{6}T_b+\frac{c}{16\pi C}+\frac{q^{2}}{4\pi K}\right)\delta T
-\frac{q^{2}}{2\pi}\mu_b\,\delta\mu.
\end{equation}
Using \eqref{eq:deltamu-linear-33}, the $\delta\mu$-dependent terms cancel exactly, and the
temperature and chemical potential decouple already at linear order:
\begin{equation}
\dot{\delta T}=-\Gamma_T\,\delta T,
\qquad
\Gamma_T=
\frac{\frac{\pi c}{6}T_b+\Lambda}{4\pi^{2}CT_b+2},
\qquad
\Lambda\equiv \frac{c}{16\pi C}+\frac{q^{2}}{4\pi K}.
\end{equation}
Therefore the late-time behavior are given by
\begin{equation}
\delta T(t)\propto e^{-\Gamma_T t},
\qquad
\delta\mu(t)\propto e^{-t/\tau_Q}.
\end{equation}
The formula for $\Gamma_T$ makes the two quantum effects explicit.  The
coefficient $\Lambda$ comes from the new linear dissipation channels in the
energy flux, while the extra $+2$ in the denominator comes from the
linear-$T$ correction to the black hole energy.  Hence the late-time
thermal relaxation is not controlled only by the classical quadratic channel; it is modified both by the quantum-corrected flux and by the quantum-corrected heat capacity.

The discussion above concerns the asymptotic approach to the bath fixed point. There is
another effect inherited from the charged classical dynamics, namely the transient
charge-driven growth discussed in subsection \ref{matterfluxes}. This effect is already
present at the classical level. Quantum corrections do not create it, but they shift the
condition under which it occurs.

\begin{figure}[t]
    \centering
    \includegraphics[width=\textwidth]{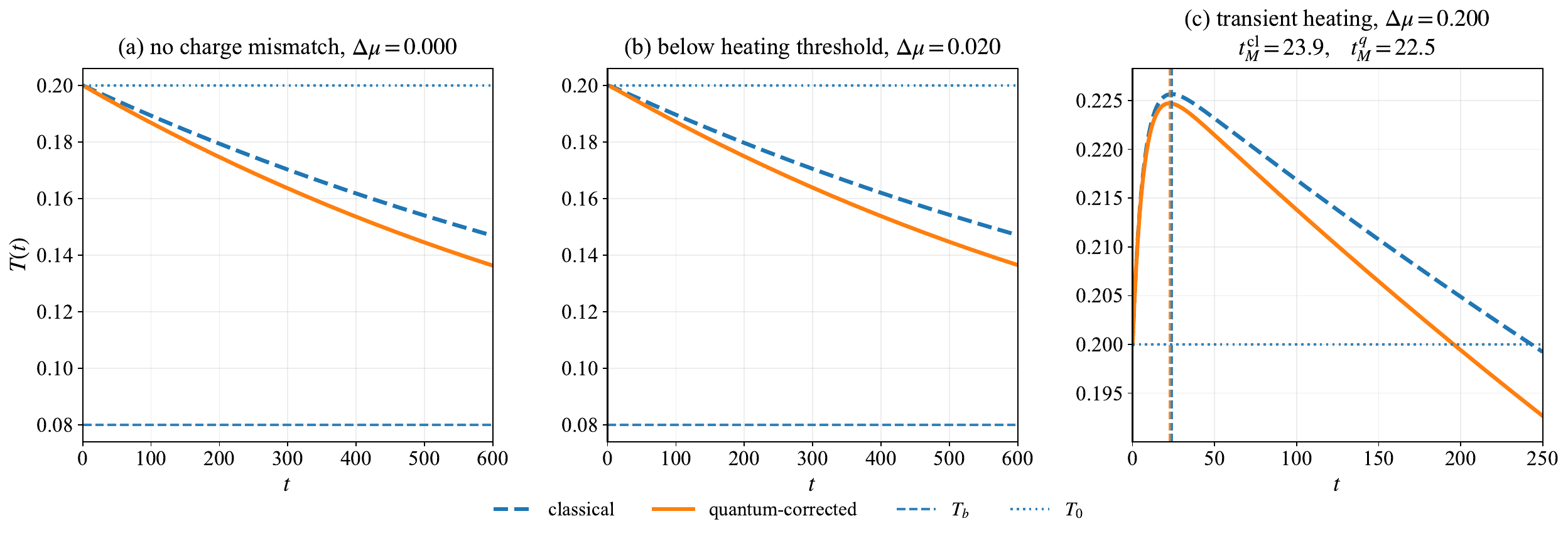}
    \caption{
    Classical and quantum-corrected temperature evolution in three representative regimes.  The parameters are
    $C=20$, $K=500$, $c=2$, $q=15$, $T_b=0.08$,
    $\mu_b=0.10$, and $T_0=0.20$.  Panel (a) shows the case
    with no chemical potential mismatch, $\Delta\mu=0$, where the charge
    source is absent and the evolution is monotonic cooling.  Panel (b) shows a
    small nonzero mismatch below the heating threshold, so that the temperature
    still decreases monotonically.  Panel (c) shows a larger mismatch above the
    threshold, for which the charge sector produces a transient heating stage.
    For this choice of parameters, the quantum-corrected linear dissipation shifts
    the turning time from $t_M^{\rm cl}\simeq 23.9$ to
    $t_M^{q}\simeq 22.5$.
    }
    \label{fig:classical-quantum-transient}
\end{figure}

\begin{figure}[t]
    \centering
    \includegraphics[width=\textwidth]{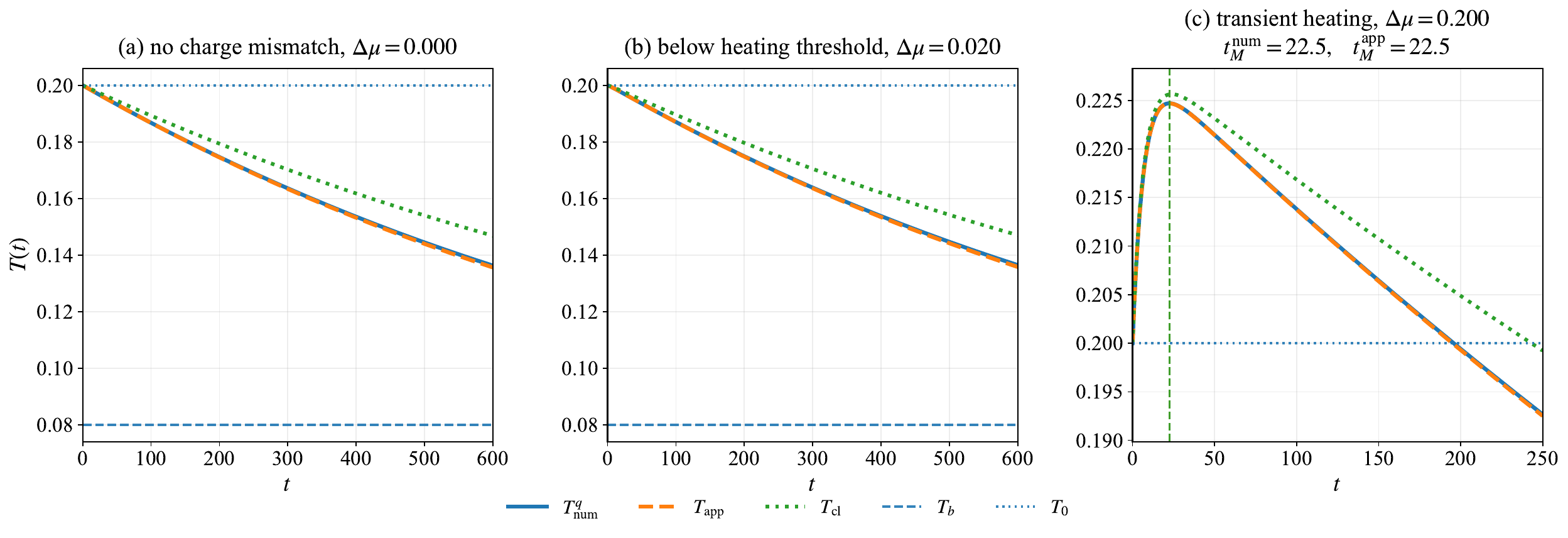}
    \caption{
    Controlled comparison between the full numerical solution and the first-order
    approximate solution of the quantum-corrected temperature evolution.  The
    parameters are the same as in figure \ref{fig:classical-quantum-transient}.  The blue curve denotes the full
    numerical solution $T^{q}_{\rm num}$, the orange dashed curve denotes
    the approximate solution $T_{\rm app}$, and the green dotted curve denotes
    the classical solution $T_{\rm cl}$.  The agreement between
    $T^{q}_{\rm num}$ and $T_{\rm app}$ shows that the controlled expansion
    captures both the monotonic cooling regimes and the transient heating regime.
    In panel (c), the turning time is reproduced as
    $t_M^{\rm num}\simeq t_M^{\rm app}\simeq 25.3$.
    }
    \label{fig:controlled-validation}
\end{figure}

After eliminating $\dot\mu$, the quantum-corrected temperature equation becomes
\begin{equation}
(4\pi^{2}CT+2)\dot T
=
-\frac{\pi c}{12}(T^2-T_b^2)
-\Lambda(T-T_b)
+\frac{q^2}{4\pi}(\mu-\mu_b)^2,
\qquad
\Lambda\equiv \frac{c}{16\pi C}+\frac{q^2}{4\pi K}.
\label{eq:q-transient heating}
\end{equation}
The last term is the same positive charge-mismatch source as in the classical theory,
whereas the term proportional to $\Lambda$ is a new quantum  linear dissipation channel. For the black hole evaporation setup $T_0>T_b$, the initial temperature increases only if
\begin{equation}
\frac{q^2}{4\pi}(\mu_0-\mu_b)^2
>
\frac{\pi c}{12}(T_0^2-T_b^2)
+\Lambda(T_0-T_b).
\label{eq:q-transient heating-condition}
\end{equation}
Compared with the classical condition, the linear dissipation term raises the threshold for
transient heating. If this condition is satisfied, the quantum-corrected turning time $t_M^q$ is defined by
\begin{equation}
\frac{q^2}{4\pi}\left(\mu(t_M^q)-\mu_b\right)^2
=
\frac{\pi c}{12}\left[T(t_M^q)^2-T_b^2\right]
+\Lambda\left[T(t_M^q)-T_b\right].
\label{eq:q-turning-time-implicit}
\end{equation}
The above argument separates three different regimes. If the chemical potential mismatch vanishes, the charge source in \eqref{eq:q-transient heating} is absent and the
black hole simply cools toward the bath.  If the mismatch is nonzero but still below
the threshold \eqref{eq:q-transient heating-condition}, the temperature again decreases monotonically, although the
quantum-corrected temperature trajectory is shifted relative to the classical one.  Finally, if the
mismatch is large enough to satisfy \eqref{eq:q-transient heating-condition}, the temperature first increases
and then it relaxes to the bath later.  These three possibilities are illustrated in
figure~\ref{fig:classical-quantum-transient}.
The purpose of figure \ref{fig:classical-quantum-transient} is only to display the
physical regimes of the full quantum-corrected equation.  For the full quantum-corrected equation, the turning condition \eqref{eq:q-turning-time-implicit} usually determines $t_M^q$ only implicitly.  To make this
condition useful, and to compare the analytic picture with the numerical
evolution, we use the controlled approximation derived in appendix \ref{appA}.  In that expansion the temperature profile
is written as
\begin{equation}
T_{\rm app}(t)
    =
    T_{\rm cl}(t)
    + \frac{1}{C}\,\delta T_C(t)
    + \frac{1}{K}\,\delta T_K(t)
    + \mathcal{O}(C^{-2},K^{-2},C^{-1}K^{-1})\,.   
\end{equation}
Here $T_{\rm cl}(t)$ is the classical solution, while
$\delta T_C(t)$ and $\delta T_K(t)$ are respectively the Schwarzian and
$U(1)$ soft-mode corrections.  The explicit single-integral expressions for
these two functions are collected in appendix \ref{appA}.
The comparison with the full numerical solution is shown in figure \ref{fig:controlled-validation}.
It is obvious that figure \ref{fig:controlled-validation} provides a consistent check that the controlled perturbative solution in the later Page time analysis is valid in the parameter range considered here.
\begin{figure}
    \centering
    \includegraphics[width=0.75\linewidth]{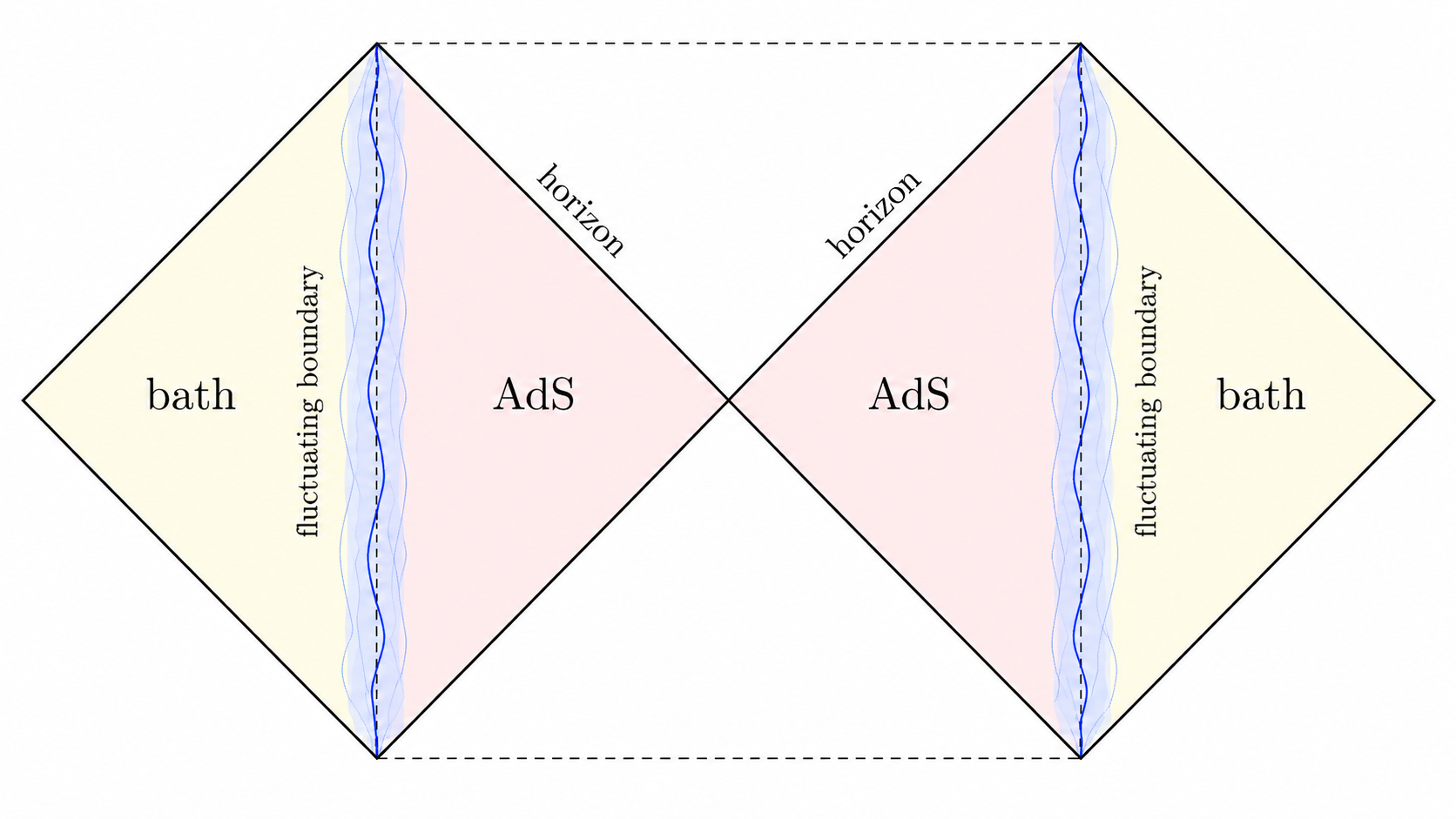}
    \caption{
Schematic geometry used in the entropy computation. 
We consider a two-sided AdS$_2$ black hole coupled to non-gravitating baths. 
The dashed vertical curves show the classical boundary saddle $F_{\rm cl}(t)$, and the solid blue curves show the corresponding quantum-corrected effective boundary trajectory $F_{\rm eff}(t)$. 
The surrounding blue bands indicate the soft-mode fluctuations of the two boundaries. 
The same effective boundary theory is placed on the left and right asymptotic boundaries.
}
    \label{schematic}
\end{figure}

\section{Page curve on the quantum-corrected evaporating background}
\label{sec:page-curve}
Here we study the Page curve on the quantum-corrected evaporating background. We do not modify the island prescription itself. What we want to know is how the corrected real time evaporation law changes the competition between the no-island saddle and the island saddle. We therefore work in the standard black hole+bath setup, shown schematically in figure \ref{schematic}, and follow the conventions of \cite{Hollowood:2020cou} adapted to the present time-dependent background.

There is one subtle point before writing the entropy formulas. The balance equation \eqref{qcbe} gives a raw quantum-averaged Schwarzian-to-temperature map
\begin{equation}
\{F_{\rm raw}(t),t\}=-2\pi^2T(t)^2-\frac{3}{2C}T(t)\,.
\end{equation}
If this raw map were used directly in the entropy formula, the late-time bath fixed point would not reduce to the standard thermal map in the desired way. For this reason we define an effective boundary trajectory $\hat F(t)$ by subtracting the static bath fixed-point shift from the one-loop Schwarzian part\,,
\begin{equation}
\{\hat F(t),t\}
=
-2\pi^2T(t)^2-\frac{3}{2C}\bigl(T(t)-T_b\bigr)\,.
\label{4.1anch}
\end{equation}
With this definition, when the system relaxes to the bath, the right-hand side of
\eqref{4.1anch} reduces to the standard thermal Schwarzian $-2\pi^2T_b^2$. 

\subsection{Entropy saddles on the quantum-corrected background}
\label{entropyonqc}
We now evaluate the no-island and island saddles on the
this quantum-corrected evaporating background. The
island prescription for evaporating two-dimensional black holes has been studied in many works \cite{Chen:2020jvn, Almheiri:2019psf, Almheiri:2019hni, Almheiri:2019yqk, Gautason:2020tmk, Hollowood:2020cou}, so we will use it here as a probe of the corrected evaporation dynamics rather than derive it again. The matter part of the entropy is computed from the standard CFT interval entropy and its conformal map generalizations \cite{Holzhey:1994we,Calabrese:2004eu}.

We restrict to the left-right symmetric setup. We keep only the right boundary endpoint and the right quantum extremal surface (QES) point explicitly. The two baths have the same inverse temperature $\beta_b$ and chemical potential $\mu_b$, and the two AdS$_2$ boundaries are described by the same effective boundary theory. Therefore there is only one independent time-dependent map, which we denote by $\hat F(t)$, or equivalently one outgoing Schwarzschild time $\hat \eta(t)$.   The endpoints of the radiation region are taken to lie at the interface between the AdS$_2$ region and the non-gravitating baths.  

The AdS$_2$ Poincare coordinates are related to the boundary null
coordinates by
\begin{equation}
  X^+=\hat F(u),\qquad X^-=\hat F(v)\,.
\end{equation}
To evaluate the CFT entropy, it is useful to introduce Schwarzschild
coordinates $y^\pm$ and the corresponding vacuum coordinates
\begin{equation}
  w^+ = e^{2\pi y^+/\beta_b},
  \qquad
  w^- = - e^{-2\pi y^-/\beta_b}.
\end{equation}
The vacuum coordinates are used to write the CFT correlators in the vacuum sate. The incoming coordinate is fixed by the boundary time,
\begin{equation}
  y^+ = t .
\end{equation}
The outgoing coordinate is related to the AdS$_2$ Poincar\'e coordinate by the following transformation
\begin{equation}
  w^- =
  \frac{\pi X^- - \beta_b}{\pi X^-+\beta_b}.
\end{equation}
With this normalization, One can check that the thermal equilibrium  reparametrization
\begin{equation}
  F_b(t)=\frac{\beta_b}{\pi}
  \tanh\left(\frac{\pi t}{\beta_b}\right)
\end{equation}
is mapped to $y^-=t$.  For the evaporating background we define
$\hat\eta(t)$ by evaluating this map on the effective boundary
trajectory,
\begin{equation}
-e^{-2\pi \hat\eta(t)/\beta_b}
=
\frac{\pi \hat F(t)-\beta_b}{\pi \hat F(t)+\beta_b},
\qquad
\hat F(t)=\frac{\beta_b}{\pi}\tanh\!\left(\frac{\pi \hat\eta(t)}{\beta_b}\right).
\end{equation}
Thus all the time dependence of the quantum-corrected evaporating
background is encoded in $\hat\eta(t)$ or $\hat F(t)$.

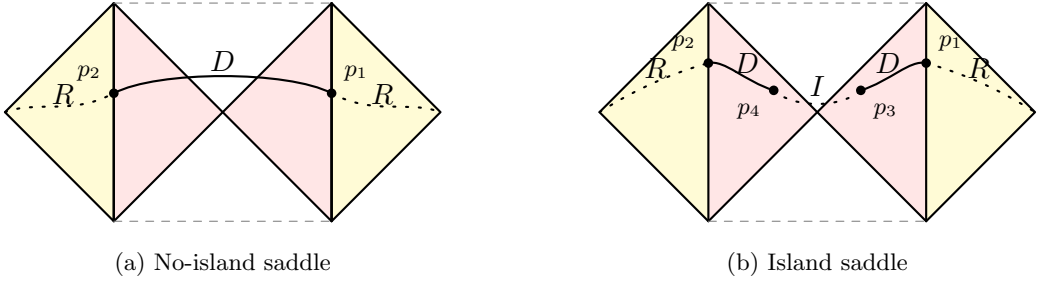
\begin{figure}[t]
\centering

\begin{minipage}[t]{0.48\textwidth}
\centering
\begin{tikzpicture}[
    x=0.48cm,y=0.48cm,
    line cap=round,
    line join=round,
    font=\footnotesize
]

\colorlet{leftbath}{gray!14}
\colorlet{rightads}{red!10}
\colorlet{rightbath}{yellow!20}

\coordinate (L)  at (-6,0);
\coordinate (LT) at (-3,3);
\coordinate (LB) at (-3,-3);
\coordinate (C)  at (0,0);
\coordinate (RT) at (3,3);
\coordinate (RB) at (3,-3);
\coordinate (R)  at (6,0);

\fill[rightbath]  (L) -- (LT) -- (LB) -- cycle;
\fill[rightads]   (LT) -- (C)  -- (LB) -- cycle;
\fill[rightads]   (RT) -- (C)  -- (RB) -- cycle;
\fill[rightbath]  (RT) -- (R)  -- (RB) -- cycle;

\draw[thick] (L) -- (LT) -- (LB) -- cycle;
\draw[thick] (LT) -- (C)  -- (LB) -- cycle;
\draw[thick] (RT) -- (C)  -- (RB) -- cycle;
\draw[thick] (RT) -- (R)  -- (RB) -- cycle;

\draw[thick] (LT) -- (LB);
\draw[thick] (RT) -- (RB);

\draw[dashed,gray] (LT) -- (RT);
\draw[dashed,gray] (LB) -- (RB);

\coordinate (p2) at (-3,0.52);
\coordinate (p1) at ( 3,0.52);

\fill (p2) circle (1.8pt);
\fill (p1) circle (1.8pt);

\node[above left=1pt]  at (p2) {$p_2$};
\node[above right=1pt] at (p1) {$p_1$};


\draw[loosely dotted, thick]
    (L)
    .. controls (-5.10,0.34) and (-4.20,-0.11) ..
    (p2);

\draw[thick]
    (p2)
    .. controls (-1.80,1.15) and (1.80,1.15) ..
    (p1);

\draw[loosely dotted, thick]
    (p1)
    .. controls (4.20,-0.11) and (5.10,0.34) ..
    (R);

\node at (-4.4,0.5) {\normalsize $R$};
\node at (0,1.4) {\normalsize $D$};
\node at ( 4.4,0.5) {\normalsize $R$};

\end{tikzpicture}

\vspace{1mm}
{\footnotesize (a) No-island saddle}
\end{minipage}
\hfill
\begin{minipage}[t]{0.48\textwidth}
\centering
\begin{tikzpicture}[
    x=0.48cm,y=0.48cm,
    line join=round,
    line cap=round,
    font=\footnotesize
]

\colorlet{leftbath}{gray!14}
\colorlet{rightads}{red!10}
\colorlet{rightbath}{yellow!20}

\coordinate (L)  at (-6,0);
\coordinate (TL) at (-3,3);
\coordinate (BL) at (-3,-3);
\coordinate (C)  at (0,0);
\coordinate (TR) at (3,3);
\coordinate (BR) at (3,-3);
\coordinate (R)  at (6,0);

\fill[rightbath] (L)--(TL)--(BL)--cycle;
\fill[rightads]  (TL)--(C)--(BL)--cycle;
\fill[rightads]  (TR)--(C)--(BR)--cycle;
\fill[rightbath] (R)--(TR)--(BR)--cycle;

\draw[thick] (L)--(TL)--(C)--(BL)--cycle;
\draw[thick] (C)--(TR)--(R)--(BR)--cycle;
\draw[thick] (TL)--(BL);
\draw[thick] (TR)--(BR);

\draw[dashed, color=gray] (TL)--(TR);
\draw[dashed, color=gray] (BL)--(BR);

\coordinate (p4) at (-3,1.35);
\coordinate (p3) at (-1.2,0.60);
\coordinate (p1) at ( 1.2,0.60);
\coordinate (p2) at ( 3,1.35);

\draw[loosely dotted, thick]
    (L) .. controls (-4.9,0.70) and (-3.9,1.10) .. (p4);

\draw[loosely dotted, thick]
    (p2) .. controls (3.9,1.10) and (4.9,0.70) .. (R);

\draw[thick]
    (p4) .. controls (-2.55,1.45) and (-2.00,0.85) .. (p3);

\draw[thick]
    (p1) .. controls (2.00,0.85) and (2.55,1.45) .. (p2);

\draw[loosely dotted, thick]
    (p3) .. controls (-0.65,0.10) and (0.65,0.10) .. (p1);

\fill (p4) circle (1.8pt);
\fill (p3) circle (1.8pt);
\fill (p1) circle (1.8pt);
\fill (p2) circle (1.8pt);

\node[above left=1pt]  at (p4) {$p_2$};
\node[below left=1pt]  at (p3) {$p_4$};
\node[below right=1pt] at (p1) {$p_3$};
\node[above right=1pt] at (p2) {$p_1$};

\node at (-4.45,1.2) {\normalsize $R$};
\node at ( 4.45,1.2) {\normalsize $R$};

\node at (-1.95,1.35) {\normalsize $D$};
\node at ( 1.95,1.35) {\normalsize $D$};

\node at (0,0.72) {\normalsize $I$};

\end{tikzpicture}

\vspace{1mm}
{\footnotesize (b) Island saddle}
\end{minipage}

\caption{
Entropy saddles for the radiation entropy. 
(a) In the no-island saddle, the radiation region consists of the bath intervals $R$ on the two sides, and the entropy is computed from the CFT entanglement entropy of the interval $D$ across the AdS region between the boundary points $p_2$ and $p_1$. 
(b) In the island saddle, the generalized entropy receives a dilaton contribution from the QES points $p_3$ and $p_4$, together with the CFT entropy of the disjoint intervals indicated in the figure; the boundary endpoints are still denoted by $p_1$ and $p_2$.
}
\end{figure}

Let $p_1$ denote the right boundary endpoint of the radiation region.
In the vacuum coordinates it is
\begin{equation}
  w_{p_1}^+ = e^{2\pi t/\beta_b},
  \qquad
  w_{p_1}^- = -e^{-2\pi\eta(t)/\beta_b}.
  \label{eq:right-boundary-endpoint}
\end{equation}
The left boundary endpoint $p_2$ is obtained by exchanging the two vacuum coordinates,
\begin{equation}
  w_{p_2}^+ = w_{p_1}^-,
  \qquad
  w_{p_2}^- = w_{p_1}^+ .
\label{eq:left-boundary-endpoint}
\end{equation}
Then the entropy of radiation region $R$ is given by
\begin{equation}
  S_{\rm CFT}(R)=S_{\rm CFT}(D)
  =
  \frac{c}{6}
  \log\left[
  -\frac{
  (w_{p_1}^+-w_{p_2}^+)(w_{p_1}^- - w_{p_2}^-)
  }{
  \Omega_{p_1}\Omega_{p_2}
  }
  \right] .
\end{equation}
Here $\Omega_{p_1}$ and $\Omega_{p_2}$ are the conformal factors at the two
endpoints in the $w^\pm$ frame. 
The conformal factor at either boundary endpoint is
\begin{equation}
  \Omega_{p_i}
  =
  \left|
  \frac{dw_{p_i}^+}{dt}
  \frac{dw_{p_i}^-}{dt}
  \right|^{1/2}
  =
  \frac{2\pi}{\beta_b}
  \exp\left[
  \frac{\pi}{\beta_b}\bigl(t-\eta(t)\bigr)
  \right]
  \sqrt{\eta'(t)} .
\end{equation}

The no-island branch is given by the CFT entropy of the interval connecting $p_1$ and $p_2$. This gives
\begin{equation}
  S_{\rm no}(t)
  \equiv S_{\rm CFT}(D)=
  \frac{c}{3}\log\!\left[
\frac{\beta_b}{\pi}
\cosh\!\left(\frac{\pi}{\beta_b}\bigl(t+\hat\eta(t)\bigr)\right)
\right]
-\frac{c}{6}\log \hat\eta'(t).
\end{equation}
This is the standard no-island result evaluated on the effective
evaporating background. We next turn to the island saddle.

Let
\begin{equation}
  p_3=(u_a,v_a)
\end{equation}
be the right QES point in the gravitating region.  Its left reflected
image is denoted by $p_4$ and will not be written explicitly.  The dilaton entering the geometric term should be evaluated on the same effective background as the no island entropy.  Before defining the effective
dilation from the effective boundary reparametrization $F_{\rm eff}$, let us first explain how to construct the bulk dilation field for a fixed boundary reparametrization $F$.

As reviewed in subsection \ref{scbhsatd}, for the static saddle we have
\begin{equation}
F_0(t)=\frac{\beta}{\pi}\tanh\!\left(\frac{\pi t}{\beta}\right),
\qquad
X^+=F_0(u),\quad X^-=F_0(v),    
\end{equation}
and the corresponding dilaton profile can be written in Poincar\'e coordinates as
\begin{equation}
\Phi(X^+,X^-)
=
2\phi_r\,
\frac{1-(\pi T)^2X^+X^-}{X^+-X^-},
\qquad T=\beta^{-1}.    
\end{equation}
Writing this in the $(u,v)$ frame gives
\begin{equation}
\Phi_{F_0}(u,v)
=
2\phi_r\,
\frac{1-(\pi T)^2F_0(u)F_0(v)}{F_0(u)-F_0(v)}.    
\end{equation}
Now use the following identities
\begin{equation}
F_0'(x)=1-(\pi T)^2F_0(x)^2,
\qquad
\frac12\frac{F_0''(x)}{F_0'(x)}=-(\pi T)^2F_0(x).    
\end{equation}
Then
\begin{equation}
\begin{aligned}
\frac12\frac{F_0''(v)}{F_0'(v)}
+\frac{F_0'(v)}{F_0(u)-F_0(v)}
&=
-(\pi T)^2F_0(v)
+\frac{1-(\pi T)^2F_0(v)^2}{F_0(u)-F_0(v)} \\
&=
\frac{1-(\pi T)^2F_0(u)F_0(v)}{F_0(u)-F_0(v)} .
\end{aligned}    
\end{equation}
This motivates the following bulk reconstruction formula
\begin{equation}
\Phi_F(u,v)
=
2\phi_r
\left[
\frac12\frac{F''(v)}{F'(v)}
+
\frac{F'(v)}{F(u)-F(v)}
\right] .
\label{eq:reconstruct}
\end{equation}
In the low-temperature regime considered here, we use the same reconstruction formula for the effective dilaton, with the classical boundary trajectory replaced by the effective one.
\begin{equation}
\Phi_{\hat F}(u,v)
=
2\phi_r\left[
\frac12\frac{\hat F''(v)}{\hat F'(v)}
+
\frac{\hat F'(v)}{\hat F(u)-\hat F(v)}
\right].
\end{equation}
The role of the $U(1)$ sector is to modify the boundary thermodynamic relation and hence the $F_{\rm eff}$, rather than to change the relation \eqref{eq:reconstruct} between $\Phi$ and $F$ itself.

In the late-time factorization channel, the island saddle which involves four-point twist
correlator reduces to two identical boundary--QES two-point
contributions.  Using the left-right symmetry, the
island entropy can be written only as a one-sided expression multiplied twice
\begin{equation}
S_{\rm isl}(t)
=
\underset{p_3}{\rm Ext}
\left[
\frac{\Phi_0+\Phi_{\hat F}(p_3)}{2G_2}
+
\frac{c}{3}
\log\!\left(
-\frac{(w^+_{p_1}-w^+_{p_3})(w^-_{p_1}-w^-_{p_3})}
{\Omega_{p_1}\Omega_{p_3}}
\right)
\right].
\label{eq:island}
\end{equation}
The QES position is determined by
\begin{equation}
  \partial_{u_a} S_{\rm isl}=0,
  \qquad
  \partial_{v_a} S_{\rm isl}=0 .
\end{equation}
The left QES equation is the reflected copy of these two conditions. In evaluating the
matter four-point function, we keep only the leading factorized channel
$(p_1,p_3)\,,(p_2,p_4)$, while the crossed contribution is neglected.\footnote{A concrete
estimate follows from the vacuum coordinates of the boundary endpoints \eqref{eq:right-boundary-endpoint}, \eqref{eq:left-boundary-endpoint} and the QES position \eqref{eq:QES}. Writing
$a\equiv \beta_b k/(2\pi)\ll1$, the factorized pairing contains
$w_{13}^+=(1-a)e^{2\pi t/\beta_b}$ and
$w_{13}^-=-(1-a)e^{-2\pi\eta_0(t)/\beta_b}$,
whereas the crossed pairing contains
$w_{14}^+=e^{2\pi t/\beta_b}+a e^{-2\pi\eta_0(t)/\beta_b}$ and
$w_{14}^-=-e^{-2\pi\eta_0(t)/\beta_b}-a e^{2\pi t/\beta_b}$,
and similarly for $w_{23}^{\pm}$. In the regime
$t\sim \eta_0(t)\sim k^{-1}$ with $\beta_b k\ll1$, one has
$e^{2\pi t/\beta_b}\gg1$ and $e^{-2\pi\eta_0(t)/\beta_b}\ll1$, so the crossed distances
are dominated by the large exponential pieces. The ratio between the crossed and factorized
channels is then parametrically of order
$\exp[-\mathcal{O}(1/(\beta_b k))]$. Hence the crossed term is nonperturbatively small and lies
beyond the $1/C$ and $1/K$ linear order kept in the present analysis.}
In other words, within the late-time semiclassical window relevant here, the crossed
channel is parametrically suppressed and does not affect the leading-order Page time shift
computed below.

With this approximation, and still in the working semiclassical low-temperature regime \eqref{semiclassical_low}, the QES equation can be solved analytically. This solution will be used as the
zeroth-order reference point for the perturbative analysis below. We therefore set
\begin{equation}
F_{\rm eff}\to F_0,\qquad \eta_{\rm eff}\to \eta_0.
\end{equation}

where $F_0$ is the classical trajectory and $\eta_0$ is the
corresponding outgoing Schwarzschild time.  When the black hole+bath is under the equilibrium, we have $\eta_0(t)=t$.  In the semiclassical regime
\begin{equation}
  \beta_b k\ll 1,
  \qquad
  k\equiv\frac{c}{24\pi C},
\end{equation}
Recall that $k$ is the evaporation rate and $c$ is $\mathcal{O}(1)$ central charge. The QES equations can be solved analytically.  The right QES is located
at
\begin{equation}
  w^+_{p_3}
  =
  \frac{\beta_b k}{2\pi}
  e^{2\pi t/\beta_b},
  \qquad
  w^-_{p_3}
  =
  -\frac{\beta_b k}{2\pi}
  e^{-2\pi\eta_0(t)/\beta_b}.
  \label{eq:QES}
\end{equation}
Substituting the above solution into the island entropy formula \eqref{eq:island} gives the classical island branch
\begin{equation}
  S^{(0)}_{\rm isl}(t;\eta_0)
  =
  \frac{\Phi_0}{2G_2}
  +
  \frac{8\pi^2 C}{\beta_b}
  +
  \frac{c}{3}\log\frac{\beta_b}{\pi}
  +
  \frac{\pi c}{3\beta_b}
  \bigl(t-\eta_0(t)\bigr)
  -
  \frac{c}{6}\log \eta_0'(t).
\end{equation}
In the equilibrium limit $\eta_0(t)=t$, the island saddle becomes the constant
plateau
\begin{equation}
  S^{(0)}_{\rm isl}
  =
  \frac{\Phi_0}{2G_2}
  +
  \frac{8\pi^2 C}{\beta_b}
  +
  \frac{c}{3}\log\frac{\beta_b}{\pi}.
\end{equation}
The corresponding zeroth-order no-island branch is
\begin{equation}
  S^{(0)}_{\rm no}(t;\eta_0)
  =
  \frac{c}{3}
  \log\left[
  \frac{\beta_b}{\pi}
  \cosh\left(
  \frac{\pi}{\beta_b}
  \bigl(t+\eta_0(t)\bigr)
  \right)
  \right]
  -
  \frac{c}{6}\log\eta_0'(t).
\end{equation}
The zeroth-order Page time $t_0\equiv t_{\rm Page}^{(0)}$ is therefore
determined by
\begin{equation}
  S^{(0)}_{\rm no}(t_0;\eta_0)
  =
  S^{(0)}_{\rm isl}(t_0;\eta_0).
\end{equation}

The quantum-corrected computation keeps the same structure, but replaces the zeroth-order map $\eta_0$ and the classical dilaton $\Phi_{F_0}$ by the effective quantities $\eta_{\rm eff}$ and
$\Phi_{F_{\rm eff}}$.  The effect of the $1/C$ and $1/K$
corrections is therefore to shift the relative position of the
no-island and island branches, and hence the Page time.

\subsection{Quantum corrections to the competition between the no-island and island saddles}
\label{section:qcisland}
After obtaining the zeroth-order saddles, we ask how the corrected background changes their relative position. In the present setup, the relevant deformation is
\begin{equation}
\hat{\eta}(t)=\eta_0(t)+\delta\eta(t), \qquad \Phi_{F_0}\rightarrow \Phi_{\hat F}.
\end{equation}
The problem is then to determine how this deformation shifts the relative entropy of the two branches, and hence the Page curve and the Page time. There is, however, a simple difference between the two branches. The no-island entropy is directly determined once the background trajectory is known. The island entropy is slightly different, since it is defined after
extremizing the generalized entropy with respect to the QES position. We therefore treat the island branch and the no-island branch separately.

We first consider the island saddle. Let
\begin{equation}
x\equiv (u_a,v_a)
\end{equation}
denote the QES position, and write the generalized entropy functional as
\begin{equation}
S_{\rm isl}(t;x)=S_{\rm isl}^{(0)}(t;x)+\delta S_{\rm isl}(t;x),
\end{equation}
where $S_{\rm isl}^{(0)}$ is the zeroth-order functional and $\delta S_{\rm isl}$ is the correction
induced by the replacement $\eta_0\to\eta_{\rm eff}$ and $\Phi_{F_0}\to\Phi_{F_{\rm eff}}$.
We first expand the quantum-corrected QES position as
\begin{equation}
x_*(t)=x_0(t)+\delta x(t),
\end{equation}
where $x_0=(u_{a,0},v_{a,0})$ satisfies the zeroth-order extremality condition
\begin{equation}
\partial_x S_{\rm isl}^{(0)}(t;x_0)=0.
\end{equation}
Next, expanding the corrected extremality condition
\begin{equation}
\partial_x S_{\rm isl}(t;x_*)=0
\end{equation}
to first order gives
\begin{equation}
\partial_i\partial_j S_{\rm isl}^{(0)}(t;x)\Big|_{x_0}\,\delta x^j
+\partial_i \delta S_{\rm isl}(t;x)\Big|_{x_0}=0,
\qquad i,j\in\{u_a,v_a\}.
\end{equation}

For the on-shell island entropy, however, the first-order correction is simpler.  Expanding $S_{\rm isl}(t;x_*)$ around $x_0$, one finds
\begin{align}
S_{\rm isl}(t)
&\equiv S_{\rm isl}(t;x_*)
\nonumber\\
&=
S_{\rm isl}^{(0)}(t;x_0)
+\partial_x S_{\rm isl}^{(0)}(t;x)\Big|_{x_0}\!\cdot\delta x
+\delta S_{\rm isl}(t;x_0)
+\mathcal{O}(\delta^2).
\end{align}
Since $x_0$ is already an extremum of the zeroth-order functional, the linear term
proportional to $\partial_x S_{\rm isl}^{(0)}(t;x_0)$ vanishes. Therefore,
\begin{equation}
S_{\rm isl}(t)=S_{\rm isl}^{(0)}(t;x_0)+\delta S_{\rm isl}(t;x_0)+\mathcal{O}(\delta^2).
\end{equation}
In other words, to first order it is sufficient to evaluate the corrected functional on the
zeroth-order QES. Using the zeroth-order on-shell island entropy obtained in section \ref{entropyonqc},
\begin{equation}
S_{\rm isl}^{(0)}(t;\eta_0)
=
\frac{\Phi_0}{2G_2}
+\frac{8\pi^2 C}{\beta_b}
+\frac{c}{3}\log\frac{\beta_b}{\pi}
+\frac{\pi c}{3\beta_b}\bigl(t-\eta_0(t)\bigr)
-\frac{c}{6}\log \eta_0'(t),
\end{equation}
and expanding it under $\eta_0\rightarrow\eta_{\rm eff}=\eta_0+\delta\eta$, we obtain
\begin{equation}
\delta S_{\rm isl}(t)
=
-\frac{\pi c}{3\beta_b}\,\delta\eta(t)
-\frac{c}{6}\frac{\delta\eta'(t)}{\eta_0'(t)}.
\end{equation}
The no-island branch is simpler because there is no additional extremization. In this case the correction is obtained directly by expanding the zeroth-order expression
\begin{equation}
S_{\rm no}^{(0)}(t;\eta_0)
=
\frac{c}{3}\log\!\left[
\frac{\beta_b}{\pi}
\cosh\!\left(\frac{\pi}{\beta_b}\bigl(t+\eta_0(t)\bigr)\right)
\right]
-\frac{c}{6}\log \eta_0'(t).
\end{equation}
Defining
\begin{equation}
x_0(t)\equiv \frac{\pi}{\beta_b}\bigl(t+\eta_0(t)\bigr),
\end{equation}
one finds
\begin{equation}
\delta S_{\rm no}^{\rm (dyn)}(t)
=
\frac{\pi c}{3\beta_b}\tanh x_0(t)\,\delta\eta(t)
-\frac{c}{6}\frac{\delta\eta'(t)}{\eta_0'(t)}.
\end{equation}

Now we can compare the two branches directly. It is convenient to introduce the entropy difference
\begin{equation}
\Delta S(t)\equiv S_{\rm no}(t)-S_{\rm isl}(t),
\qquad
\Delta S^{(0)}(t)\equiv S_{\rm no}^{(0)}(t;\eta_0)-S_{\rm isl}^{(0)}(t;\eta_0).
\end{equation}
Subtracting the two branch corrections, the $\delta\eta'$-terms cancel identically:
\begin{equation}
\delta\Delta S(t)
=
\delta S_{\rm no}(t)-\delta S_{\rm isl}(t)
=
\frac{\pi c}{3\beta_b}
\left[1+\tanh x_0(t)\right]\delta\eta(t).
\end{equation}
This cancellation is useful because it shows that, at leading order, the shift of the
competition depends only on $\delta\eta(t)$ itself, and not on its derivative.

The Page time shift then follows by expanding the crossing condition around the zeroth-order Page time. Let
\begin{equation}
t_{\rm Page}=t_0+\delta t_{\rm Page},
\end{equation}
where $t_0\equiv t_{\rm Page}^{(0)}$ is the zeroth-order Page time determined by
\begin{equation}
\Delta S^{(0)}(t_0)=0.
\end{equation}
Expanding the corrected crossing condition
\begin{equation}
\Delta S(t_{\rm Page})=0
\end{equation}
to first order gives
\begin{equation}
\delta t_{\rm Page}
=
-\frac{\delta\Delta S_(t_0)}{\Delta S^{(0)\prime}(t_0)}.
\end{equation}
To make this expression more explicit, define
\begin{equation}
x_p^{(0)}\equiv x_0(t_0)=\frac{\pi}{\beta_b}\bigl(t_0+\eta_0(t_0)\bigr),
\qquad
\delta\eta_p\equiv \delta\eta(t_0).
\end{equation}
Then
\begin{equation}
\delta\Delta S(t_0)
=
\frac{\pi c}{3\beta_b}\bigl(1+\tanh x_p^{(0)}\bigr)\delta\eta_p.
\end{equation}
On the other hand, differentiating the zeroth-order entropy difference yields
\begin{equation}
\Delta S^{(0)\prime}(t_0)
=
\frac{\pi c}{3\beta_b}D_p^{(0)},
\end{equation}
with
\begin{equation}
D_p^{(0)}
\equiv
\bigl(1+\eta_0'(t_0)\bigr)\tanh x_p^{(0)}
-
\bigl(1-\eta_0'(t_0)\bigr).
\label{eq:def_Dp}
\end{equation}
Therefore, the dynamical contribution to the Page time shift can be written in the compact
form
\begin{equation}
\delta t_{\rm Page}
=
-\frac{1+\tanh x_p^{(0)}}{D_p^{(0)}}\,\delta\eta_p
\equiv
- F_p^{(0)}\,\delta\eta_p,
\label{eq:compactform}
\end{equation}
where
\begin{equation}
F_p^{(0)}
\equiv
\frac{1+\tanh x_p^{(0)}}{D_p^{(0)}}.
\end{equation}
This is the form that we will use below when we discuss separately the $1/C$ and $1/K$
contributions to the Page time shift.

At this stage it is useful to comment on the sign of the prefactor $F_p^{(0)}$.
Since
\begin{equation}
F_p^{(0)}=\frac{1+\tanh x_p^{(0)}}{D_p^{(0)}},
\end{equation}
the numerator is always positive, and hence the sign of $F_p^{(0)}$ is entirely
determined by $D_p^{(0)}$. Using the definition \eqref{eq:def_Dp}, one can rewrite
\begin{equation}
D_p^{(0)}
=
\bigl(1+\eta_0'(t_0)\bigr)\tanh x_p^{(0)}-\bigl(1-\eta_0'(t_0)\bigr)
=
\frac{2\bigl(\eta_0'(t_0)-e^{-2x_p^{(0)}}\bigr)}{1+e^{-2x_p^{(0)}}}.
\end{equation}
Therefore,
\begin{equation}
\mathrm{sign}\!\bigl(F_p^{(0)}\bigr)=\mathrm{sign}\!\bigl(D_p^{(0)}\bigr)
=\mathrm{sign}\!\Bigl(\eta_0'(t_0)-e^{-2x_p^{(0)}}\Bigr).
\end{equation}
In particular, $F_p^{(0)}$ is positive provided
\begin{equation}
\eta_0'(t_0)>e^{-2x_p^{(0)}}.
\end{equation}
At the zeroth-order Page point $t_0$, the crossing condition
$S_{no}^{(0)}(t_0;\eta_0)=S_{isl}^{(0)}(t_0;\eta_0)$ implies
\begin{equation}
\log\cosh x_p^{(0)}
=
\frac{3\Phi_0}{2c\,G_2}
+\frac{24\pi^2 C}{c\,\beta_b}
+\frac{\pi}{\beta_b}\bigl(t_0-\eta_0(t_0)\bigr).
\end{equation}
In the semiclassical regime $4\pi^2 C/\beta_b\gg1$, the second term on the right-hand
side is parametrically large. Therefore, $\log\cosh x_p^{(0)}\gg1$, which implies
$x_p^{(0)}\gg1$. This means $e^{-2x_p^{(0)}}$ is small, while $\eta_0'(t_0)$ remains parametrically larger in this semiclassical regime. Hence we have
\begin{equation}
D_p^{(0)}>0,
\qquad
F_p^{(0)}>0.
\end{equation}
This observation will be used in the sign analysis below.

To summarize, the most important result is that at first order the dynamical contribution to the Page time shift can be
written as
\begin{equation}
\delta t_{\rm Page}
=
- F_p^{(0)}\,\delta\eta_p,
\qquad
\delta\eta_p \equiv \delta\eta(t_0),
\end{equation}
where $t_0 \equiv t_{\rm Page}^{(0)}$ is the zeroth-order Page time and $F_p^{(0)}$ is
evaluated on the zeroth-order background. Therefore, once the effective map
$\eta_{\rm eff}(t)$ is known, the dynamical Page time shift follows immediately from its
value at $t=t_0$.

For the purposes of the main text, we do not need to repeat the full reconstruction of
$\eta_{\rm eff}(t)$ and $F_{\rm eff}(t)$. The only input needed for the entropy analysis is
that these quantities are determined by the quantum-corrected temperature profile $T(t)$.
More precisely, appendix \ref{appB} starts from 
\begin{equation}
\hat F(t)=\frac{\beta_b}{\pi}\tanh\!\left(\frac{\pi}{\beta_b}\hat\eta(t)\right),
\end{equation}
and combines it with the Schwarzian equation satisfied by $\hat F(t)$ to
reconstruct $\hat\eta(t)$. This yields the expansion
\begin{equation}
\hat\eta(t)
=
\eta_0(t)
+\frac{1}{C}\eta_C(t)
+\frac{1}{K}\eta_K(t)
+\mathcal{O}(C^{-2},K^{-2},C^{-1}K^{-1})\,.
\label{4.57}
\end{equation}
Substituting the expansion \eqref{4.57} into \eqref{eq:compactform}, the Page time shift decomposes into the Schwarzian and $U(1)$ channels,
\begin{equation}
\delta t_{\rm Page}
=
-\frac{F_p^{(0)}}{C}\eta_C(t_0)
-\frac{F_p^{(0)}}{K}\eta_K(t_0)
+\mathcal{O}(C^{-2},K^{-2},C^{-1}K^{-1}),
\end{equation}
where $t_0\equiv t_{\rm Page}^{(0)}$ is the zeroth-order Page time. It is therefore natural to define
\begin{equation}
\delta t_{\rm Page}^{(C)}\equiv -\frac{F_p^{(0)}}{C}\eta_C(t_0),
\qquad
\delta t_{\rm Page}^{(K)}\equiv -\frac{F_p^{(0)}}{K}\eta_K(t_0),
\end{equation}
so that
\begin{equation}
\delta t_{\rm Page}
=
\delta t_{\rm Page}^{(C)}+\delta t_{\rm Page}^{(K)}
+\mathcal{O}(C^{-2},K^{-2},C^{-1}K^{-1}).
\end{equation}
Using the equations \eqref{B.28} and \eqref{B.29} in appendix \ref{appB}, we have the following integral formulas

\begin{equation}
\eta_C(t_0)
=
\frac{1}{8\pi^2T_b}
\int_0^{t_0}ds\,
\frac{X_C(s)+3\bigl(T_{\rm cl}(s)-T_b\bigr)}{T_{\rm cl}(s)},
\qquad
\eta_K(t_0)
=
\frac{3q^2}{2\pi^2 c\,T_b}
\int_0^{t_0}ds\,
\frac{X_K(s)}{T_{\rm cl}(s)},
\end{equation}
with
\begin{equation}
X_C(t)
=
-e^{-\gamma t}\int_0^t dt'\,e^{\gamma t'}
\Bigl[
4\dot T_{\rm cl}(t')+3\gamma\bigl(T_{\rm cl}(t')-T_b\bigr)
\Bigr],
\end{equation}
and
\begin{equation}
X_K(t)
=
-e^{-\gamma t}\int_0^t dt'\,\gamma e^{\gamma t'}
\bigl[T_{\rm cl}(t')-T_b\bigr].
\end{equation}
To make the sign structure more transparent, let
\begin{equation}
\Delta(t)\equiv T_{\rm cl}(t)-T_b.
\end{equation}
Then the $1/K$ channel can be rewritten as
\begin{equation}
\eta_K(t_0)
=
-\frac{3q^2}{2\pi^2 c\,T_b}
\int_0^{t_0}du\,W_K(u)\,\Delta(u),
\qquad
W_K(u):=
e^{\gamma u}
\int_u^{t_0}dt\,
\frac{\gamma e^{-\gamma t}}{T_{\rm cl}(t)}
>0.
\end{equation}
Therefore, whenever $T_{\rm cl}(t)>T_b$ on the interval $[0,t_0]$, one has
$\eta_K(t_0)<0$, and thus
\begin{equation}
\delta t_{\rm Page}^{(K)}>0.
\end{equation}
In other words, the $1/K$ correction from the $U(1)$ phase mode tends to delay the Page transition.

For the Schwarzian channel, it is convenient to define
\begin{equation}
I_1(t_0):=
\int_0^{t_0} dt\,
\frac{T_b+e^{-\gamma t}(4T_i-4T_b)-T_{\rm cl}(t)}{T_{\rm cl}(t)},
\qquad
I_2(t_0):=
\int_0^{t_0} du\,W_K(u)\,\Delta(u).
\label{eq:I1I2}
\end{equation}
Then
\begin{equation}
\eta_C(t_0)=\frac{1}{8\pi^2 T_b}\Bigl[I_1(t_0)+I_2(t_0)\Bigr].
\end{equation}
Hence, 
\begin{equation}
\delta t_{\rm Page}^{(C)}<0
\quad\Longleftrightarrow\quad
I_1(t_0)+I_2(t_0)>0.
\label{eq:generalsign}
\end{equation}
This is the general sign criterion for the Schwarzian contribution.

If $T_{\rm cl}(t)>T_b$ on the interval $[0,t_0]$, then $\Delta(u)>0$ and therefore
$I_2(t_0)>0$. In addition, the sufficient condition
\begin{equation}
T_{\rm cl}(t)<T_b+e^{-\gamma t}(4T_i-4T_b),
\qquad 0\le t\le t_0,
\label{eq:sufficient}
\end{equation}
implies $I_1(t_0)>0$, and hence guarantees
\begin{equation}
\delta t_{\rm Page}^{(C)}<0.
\label{eq:strongsign}
\end{equation}
We stress, however, that \eqref{eq:sufficient} is only a sufficient condition, and is not implied by
monotonic cooling by itself. Therefore, for the Schwarzian channel, the general statement is the criterion \eqref{eq:generalsign}, rather than the stronger sign claim \eqref{eq:strongsign}. In the numerical
examples studied in section \ref{sec_num}, the Schwarzian contribution is indeed found to advance the
Page transition.

We thus arrive at the following analytic picture. For the $1/K$ channel, one obtains a direct
sign result whenever $T_{\rm cl}(t)>T_b$ on $[0,t_0]$, namely
\begin{equation}
\delta t_{\rm Page}^{(K)}>0.
\tag{4.80}
\end{equation}
For the Schwarzian channel, by contrast, the sign is controlled by the criterion
\eqref{eq:generalsign}. The numerical results of section \ref{sec_num} show that, in the parameter range studied
in this paper, the Schwarzian contribution is negative and hence tends to move the Page time
earlier. Therefore the sign of the total Page time shift is determined by the competition
between the two soft sectors. Decreasing $C$ strengthens the Schwarzian channel, whereas
decreasing $K$ strengthens the $U(1)$ phase mode channel.

\subsection{Numerical Page curves and Page time shifts}
\label{sec_num}
We now illustrate the Page curve and the Page time shift numerically. We now compare the analytic picture of section \ref{section:qcisland} with numerical Page curves and parameter scans. Unless stated otherwise, we just set $\Phi_0=0$ and use the following parameters
\begin{equation}
C=20\,,K=500\,,c=2\,,q=15\,,T_b=0.08\,,\mu_b=0.10\,,T_0=0.20\,.
\end{equation}
\begin{figure}[t]
    \centering
    \includegraphics[width=\textwidth]{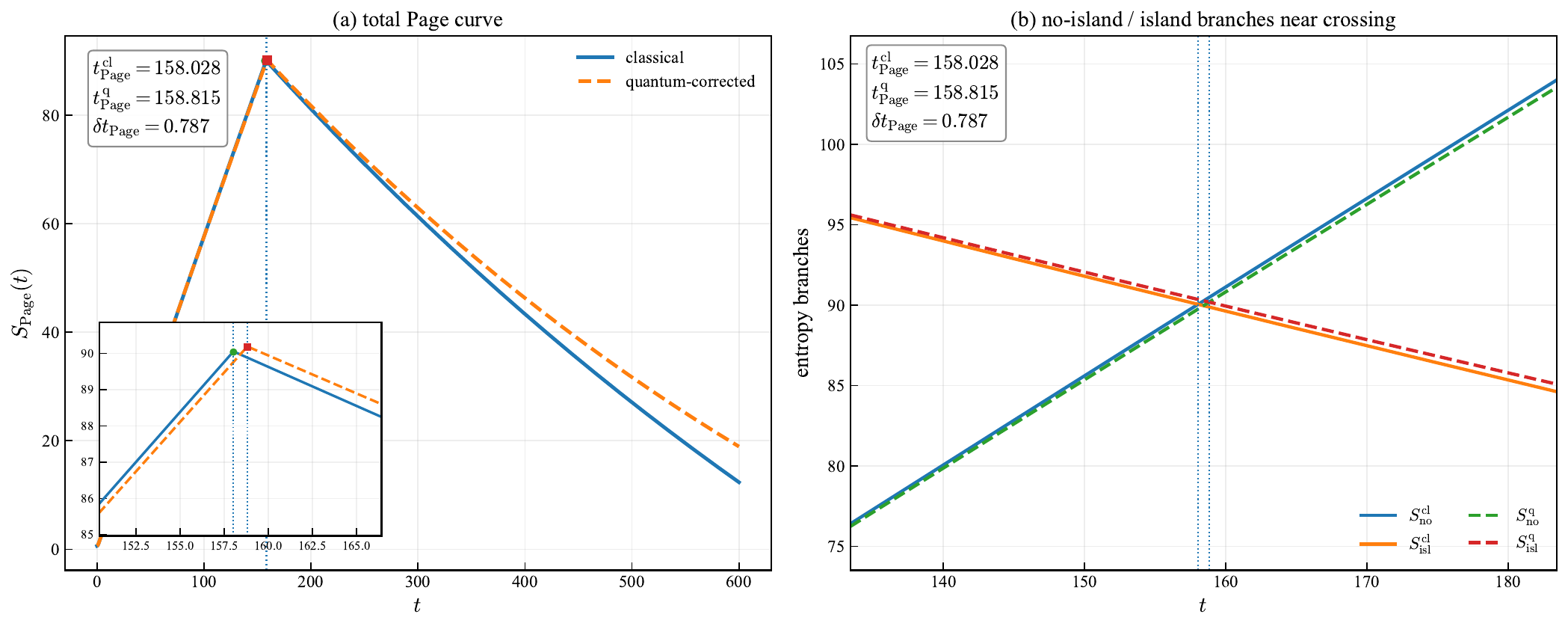}
    \caption{Page curves and entropy branches on the quantum-corrected
    background. Panel (a) shows the total Page curve $S_{\rm Page}(t)=\min\{S_{\rm no}(t),S_{\rm isl}(t)\}$
    for the classical and quantum-corrected backgrounds with the marked corresponding Page times. Panel (b) shows the no-island and island branches separately in the near crossing regime. This shows how the soft-mode quantum corrections deform the relative position of the two branches.}
    \label{fig:pagecurve_shift}
\end{figure}
Figure \ref{fig:pagecurve_shift} shows the Page curve for this fixed choice of parameters. The quantum correction produces a small but visible shift of the crossing between the no-island and island branches. The right panel shows the two branches near the crossing. This makes clear that the correction is not a uniform vertical shift of the entropy curve. Instead, it changes the relative competition between the two branches.

\begin{figure}[t]
    \centering
    \includegraphics[width=1.05\textwidth]{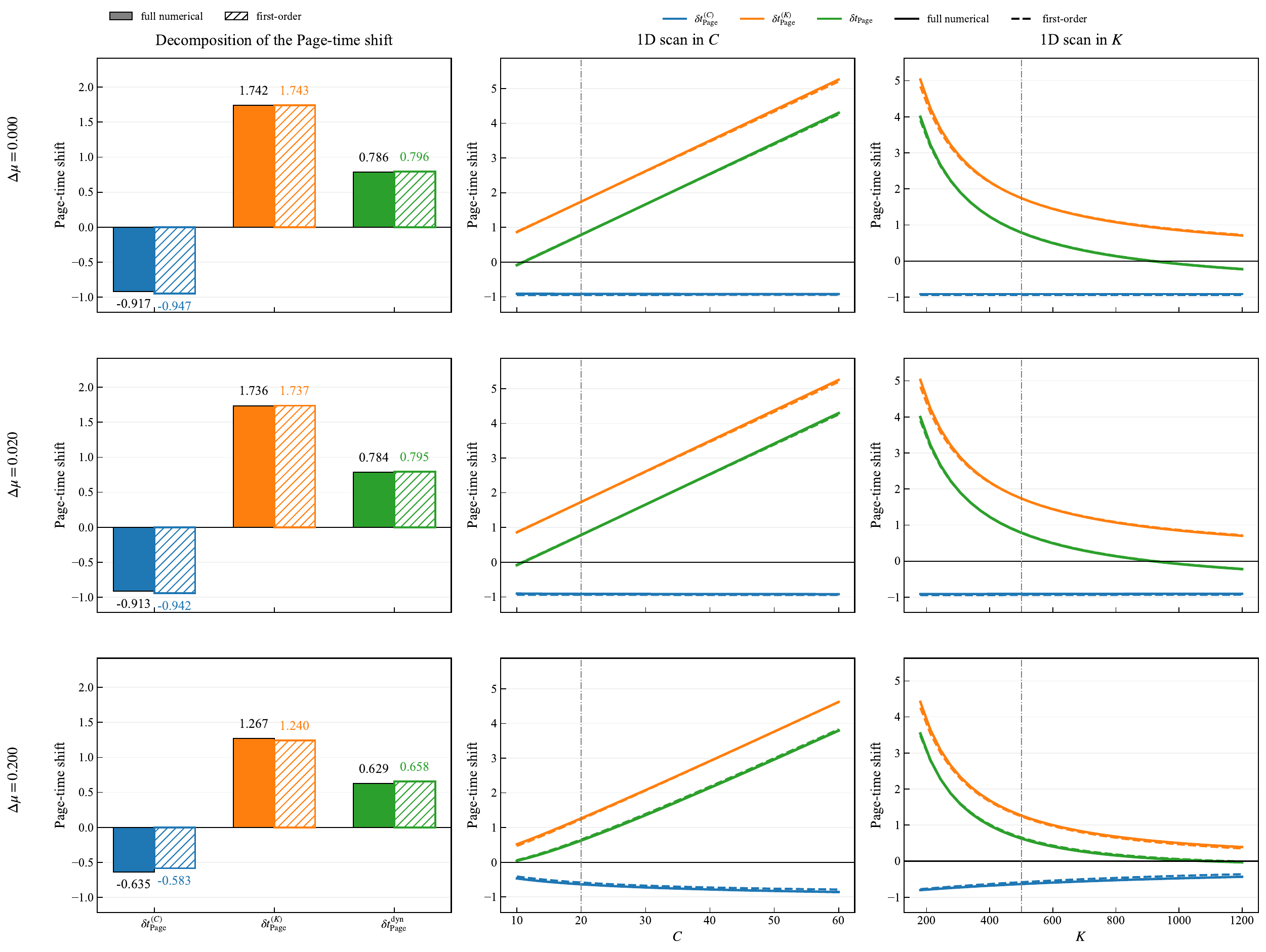}
    \caption{
Decomposition of the Page time shift and one-dimensional parameter scans. The three
rows correspond to $\Delta\mu=0$, $0.020$, and $0.200$. In the left column we use the fixed point $C=20,\ K=500$ and decompose the total shift into the Schwarzian
contribution $\delta t_{\rm Page}^{(C)}$ and the $U(1)$ contribution $\delta t_{\rm Page}^{(K)}$. The hatched bars show the first-order approximation, and the solid bars show the full numerical result. The middle and right columns show the corresponding scans in $C$ and $K$. The Schwarzian contribution tends to advance the Page transition, while the $U(1)$ contribution tends to delay it.
}
    \label{fig:decomp_1dscan}
\end{figure}

To see which part of the correction is responsible for this shift, we next separate the Page time shift into the Schwarzian and $U(1)$ contributions. Figure \ref{fig:decomp_1dscan} shows this decomposition for three values of the chemical potential mismatch,
\[
\Delta\mu=0\,,0.020\,,0.200\,.
\]
The left column uses the fixed point $C=20,\ K=500$. The middle and right columns
then vary $C$ and $K$, respectively, while keeping the other parameters fixed.

The numerical result agrees well with the first-order formula derived in section \ref{section:qcisland}. The Schwarzian contribution is negative in these examples, so it moves the Page transition to an earlier time. The $U(1)$ contribution is positive, so it delays the transition. Thus the total shift is determined by a competition between the two effects. Decreasing $C$ strengthens the Schwarzian $1/C$ effect, while decreasing $K$ strengthens the $1/K$ correction from
the $U(1)$ phase mode.

\begin{figure}[t]
\centering
\includegraphics[width=\textwidth]{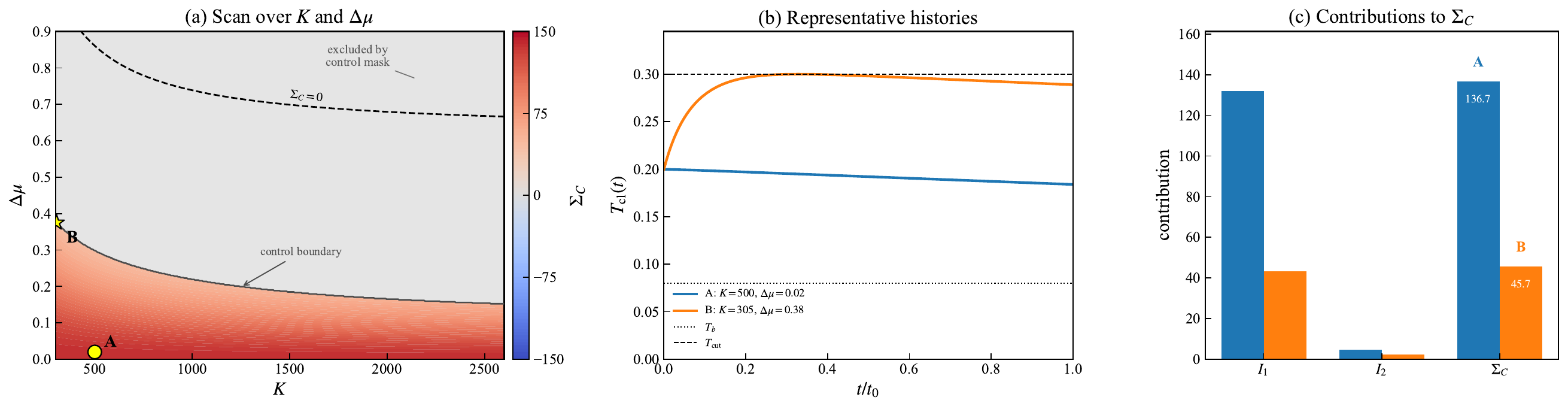}
\caption{
Sign check for the Schwarzian contribution. Panel (a) shows
$\Sigma_C=I_1+I_2$ in the $(K,\Delta\mu)$ plane. The dashed curve is the formal
contour $\Sigma_C=0$. The gray region is excluded by the near-AdS$_2$ low temperature
condition $\max_{0\leq t\leq t_0}T_{\rm cl}(t)<T_{\rm cut}$, with
$T_{\rm cut}=0.30$. The sign change of $\Sigma_C$ in this gray region is therefore
not interpreted as a controlled result. Panels (b) and (c) show two controlled points
marked in panel (a). Point A is a monotonic cooling history. Point B has a transient
heating stage but remains below $T_{\rm cut}$. Both points have $\Sigma_C>0$, so the
Schwarzian contribution advances the Page transition.
}
\label{fig:schwarzian-control-diagnostic}
\end{figure}

Before turning to the total shift on the full $(C,K)$ plane, we check the sign of the Schwarzian contribution itself. From section \ref{section:qcisland}, this sign is controlled by
\begin{equation}
\Sigma_C(t_0)\equiv I_1(t_0)+I_2(t_0),
\qquad
\delta t_{\rm Page}^{(C)}<0
\quad\Longleftrightarrow\quad
\Sigma_C(t_0)>0 \,.   
\end{equation}
Here $I_2$ is positive as long as the classical trajectory stays above the bath temperature.
The more delicate term is $I_1$. From \eqref{eq:I1I2}, $I_1$ compares the actual charged temperature profile $T_{\rm cl}(t)$ with the simple cooling profile
\[
T_b+e^{-\gamma t}(4T_i-4T_b)
\]
that appears in the Schwarzian part of the first-order correction. If $T_{\rm cl}(t)$
stays below this curve, the integrand of $I_1$ is positive. If the charge sector produces
a strong enough heating stage, $T_{\rm cl}(t)$ can rise above this curve for part of the
evolution, and that part gives a negative contribution to $I_1$.

This explains the formal $\Sigma_C=0$ curve in figure \ref{fig:schwarzian-control-diagnostic}. The charge mismatch source in the classical equation has the form
\begin{equation}
\dot X_0+\gamma(X_0-T_b^2)=\delta e^{-\lambda t},
\qquad
X_0=T_{\rm cl}^2,\qquad
\delta\propto \frac{(\Delta\mu)^2}{C},
\qquad
\lambda=\frac{q^2}{\pi K}\,.  
\end{equation}
Increasing $\Delta\mu$ increases the size of the source. Increasing $K$ makes the
charge relaxation slower. Together, these two effects produce a larger and longer-lived
heating stage. This can make $I_1$ negative enough to overcome the positive $I_2$,
so that $\Sigma_C$ becomes negative in a formal extension of the scan.

The same mechanism, however, also raises the maximum temperature. For this reason we
must impose the near-AdS$_2$ low-temperature condition
\begin{equation}
\max_{0\leq t\leq t_0}T_{\rm cl}(t)<T_{\rm cut},
\qquad
T_{\rm cut}=0.30 \,.    
\end{equation}
The gray region in figure \ref{fig:schwarzian-control-diagnostic} is the region that fails this test. It is not a new phase of the Schwarzian correction. In particular, the part of the
$\Sigma_C=0$ contour lying in the gray region should not be interpreted as a controlled
sign reversal. Inside the unmasked region shown in the figure, we find $\Sigma_C>0$,
and hence $\delta t_{\rm Page}^{(C)}<0$. Thus, within the working regime \eqref{semiclassical_low} used in this paper, we do not find a reliable regime in which simply increasing $K$ and $\Delta\mu$ changes the sign of the Schwarzian contribution.

\begin{figure}[t]
    \centering
    \includegraphics[width=1\textwidth]{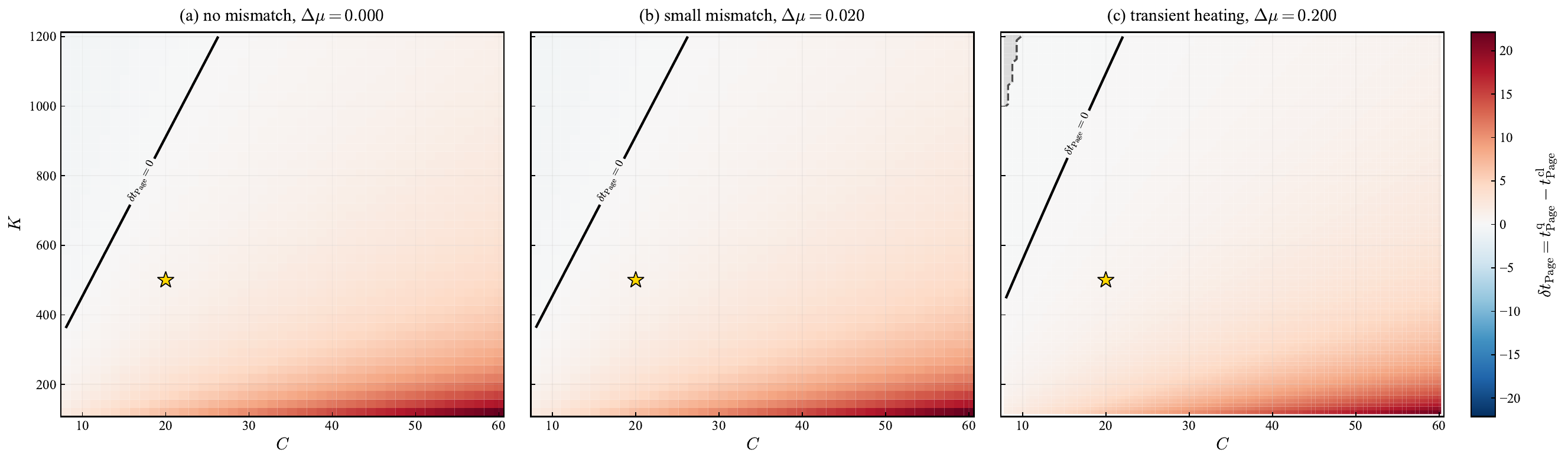}
    \caption{
Phase diagram of the Page time shift on the $(C,K)$ plane. The three panels correspond
to $\Delta\mu=0$, $0.020$, and $0.200$. The color scale gives
$\delta t_{\rm Page}=t_{\rm Page}^{q}-t_{\rm Page}^{\rm cl}$. The black curve marks
$\delta t_{\rm Page}=0$. The marked point is $C=20,\ K=500$, the fixed point used
in figures \ref{fig:pagecurve_shift} and \ref{fig:decomp_1dscan}. The gray region
is outside the near-AdS$_2$ low temperature window, $T_{\max}>T_{\rm cut}=0.30$. The negative region at small $C$ and large $K$ comes from the dominance
of the Schwarzian advance over the delay from the $U(1)$ sector.
}
    \label{fig:phase_3panel}
\end{figure}

We now turn from the Schwarzian sign check to the total Page time shift. Figure \ref{fig:schwarzian-control-diagnostic} only tested the Schwarzian contribution through
$\Sigma_C$. Figure \ref{fig:phase_3panel} shows the total shift
\begin{equation}
\delta t_{\rm Page}=t_{\rm Page}^{q}-t_{\rm Page}^{\rm cl}\,,    
\end{equation}
which includes both the Schwarzian and $U(1)$ soft sectors.

Figure \ref{fig:phase_3panel} shows the result in the $(C,K)$ plane for
three representative values of the mismatch,
\[
\Delta\mu=0,\qquad 0.020,\qquad 0.200 .
\]
For $\Delta\mu=0$, there is no charge-driven heating and the temperature decreases
monotonically from $T_0$ to $T_b$. The temperature control condition is therefore
automatically satisfied, since $T_{\max}=T_0<T_{\rm cut}$. For the small mismatch
$\Delta\mu=0.020$, the pattern is only mildly deformed.

The negative region in the upper-left part of the plots has a simple origin. It is not caused
by the sign reversal of $\Sigma_C$ discussed above. In this region the Schwarzian
contribution still advances the Page transition, while the $U(1)$ contribution delays it. Moving to smaller $C$ strengthens the negative Schwarzian contribution, since it is a
$1/C$ effect. Moving to larger $K$ weakens the positive $U(1)$ contribution, since it is
a $1/K$ effect. Therefore, for small $C$ and large $K$, the Schwarzian advance can
dominate over the $U(1)$ delay, and the total Page time shift becomes negative.

The third panel shows what changes when the mismatch is large enough to produce
transient heating. The same charge source that deforms the $\delta t_{\rm Page}=0$
contour can also push part of the evolution outside the near-AdS$_2$ temperature window.
This happens in the gray region. Those points are not used to infer the sign of the
Page time shift. Inside the unmasked region, the interpretation remains the same as in the
first two panels: the sign of the total shift is set by the competition between the
Schwarzian advance and the delay from the $U(1)$ sector.

Taken together, these figures support the analytic picture of section \ref{section:qcisland}. The quantum correction changes the Page time by changing the relative position of the no-island and island branches. The Schwarzian sector tends to move the Page transition earlier, while the $U(1)$ sector tends to move it later. The sign of the total shift is therefore not fixed by one sector alone. It is fixed by their competition, within the part of parameter space where the near-AdS$_2$ approximation remains under control.

\section{Discussion}
\label{discussion}

We studied the evaporation and Page dynamics of charged near-AdS$_2$ black holes coupled to a non-gravitating bath at fixed temperature and chemical potential. The low-energy boundary theory contains two soft sectors: the Schwarzian reparametrization
mode and the $U(1)$ phase mode. This gives a simple setting in which the real-time evaporation law and the Page curve can be studied in the same setup.

We first reviewed the charged near-AdS$_2$ boundary theory and the corresponding thermodynamic dictionary. Starting from the matter fluxes at the interface, we derived the classical balance equations for the effective temperature and chemical potential. We then included the quantum fluctuations of the two boundary soft modes by averaging the boundary energy and the outgoing flux. This led to the quantum-corrected balance equations used in the rest of the paper. Compared with the classical evolution, the corrected equations contain additional linear terms in the temperature. These terms come from the soft-mode correction to the black hole energy and from the corrected outgoing flux. They modify the relaxation of the black hole toward the bath.

We then used the corrected real-time background to compute the Page curve. The island formula itself was not changed. Instead, we asked how the corrected evaporation dynamics changes the competition between the no-island and island branches. At first order, the on-shell island entropy can be obtained by evaluating the corrected generalized entropy on the zeroth-order QES. This allows the Page time shift to be computed without solving for the corrected QES position explicitly. The shift is therefore controlled by how the corrected background changes the two entropy branches near their crossing.

The analytic result gives a clear separation between the two soft sectors. For the correction from the $U(1)$ phase mode, one obtains a direct sign result when the classical temperature stays above the bath temperature during the relevant time interval: this correction delays the Page transition. The Schwarzian correction is more subtle. Its sign is controlled by the integral criterion in equation \eqref{eq:generalsign}. In the controlled numerical region studied in this paper, this criterion gives a negative Schwarzian contribution to the Page time shift, so the Schwarzian sector tends to move the Page transition earlier. The numerical results support this picture. The Page curves show that the quantum correction changes the crossing of the no-island and island branches, rather than producing a uniform vertical shift of the entropy profile. The decomposition of the Page time shift shows good agreement between the first-order approximation and the full numerical calculation. In the examples shown before, the Schwarzian contribution advances the transition, while the $U(1)$ part delays it.

We also scanned the relevant parameter space. Decreasing $C$ strengthens the Schwarzian $1/C$ effect, while decreasing $K$ strengthens the charge-sector correction from the $U(1)$ phase mode. As a result, the total Page time shift can be positive or negative depending on which channel dominates. The negative region at small $C$ and large $K$ comes from the Schwarzian advance becoming larger than the delay from the $U(1)$ sector. For large charge mismatch, the charge sector can also produce transient heating. However, this same effect can push the evolution outside the near-AdS$_2$ low-temperature window. Those points were excluded in the phase diagram and should not be interpreted as a controlled new regime.

A related recent work studied the same Schwarzian plus $U(1)$ boundary effective theory on replica geometries and found a connected--disconnected saddle transition controlled by the temperature and the couplings $C$, $K$ \cite{Nian:2026bxm}. Our setup is different: the replica prescription is kept fixed, while the soft-mode corrections enter through the real-time evaporating background and shift the Page-time crossing.

There are several natural directions for future work. First, it would be useful to go beyond the perturbative treatment of soft-mode quantum effects used here. In this paper the Schwarzian and $U(1)$ sectors enter through the leading $1/C$ and $1/K$ corrections to the evaporation background and to the entropy branches. This is sufficient for the first-order Page-time shift, but it is not a full non-perturbative evaluation of the soft-mode path integral in the evaporating replica geometry. Such a calculation would require keeping the Schwarzian and $U(1)$ modes directly in the real-time and replica path integrals. A more microscopic treatment of the bath would also start from an influence functional, as in recent work on open JT gravity \cite{Adachi:2026hkk}. It would also be interesting to see whether the same competition between the Schwarzian sector and the charge sector appears in other near-extremal black holes.

\section*{Acknowledgements}
The author would like to thank Yuan Zhong for many helpful discussions. Special thanks to Jun Nian for his instructions throughout this work. This work is supported in part by the NSFC under grants No.~12375067 and No.~12547104.

\appendix

\section{Controlled approximation to the quantum-corrected temperature evolution}
\label{appA}
In this appendix we derive the controlled approximation to the temperature profile used in
subsection \ref{subsec:qbalance-limits} and later in appendix \ref{appB}. The final result is organized by the following equation
\begin{equation}
T(t)
=
T_{\rm cl}(t)
+\frac{1}{C}\,\delta T_C(t)
+\frac{1}{K}\,\delta T_K(t)
+\mathcal{O}\!\left((CT)^{-2},(KT)^{-2},(CT)^{-1}(KT)^{-1}\right),
\label{A.1}
\end{equation}
where $T_{\rm cl}(t)$ is the classical profile, while $\delta T_C(t)$ and $\delta T_K(t)$
denote respectively the Schwarzian and $U(1)$ first-order corrections.

Throughout this appendix we work in the same working regime as in the main text,
\begin{equation}
TL\ll 1,\qquad 4\pi^2 C T\gg 1,\qquad KT\gg 1,
\end{equation}
and we consistently keep only terms to first order in $1/(CT)$ and $1/(KT)$.

We start from the temperature equation obtained from the full quantum-corrected balance equation after eliminating $\dot\mu$,
\begin{equation}
(4\pi^2 C\,T+2)\dot T
=
-\alpha\,(T^2-T_b^2)
-\frac{c}{16\pi C}(T-T_b)
-\frac{q^2}{4\pi K}(T-T_b)
+J(t),
\label{A.3}
\end{equation}
with
\begin{equation}
\alpha\equiv \frac{\pi c}{12}.
\end{equation}
The charge equation gives
\begin{equation}
\dot\mu
=
-\frac{q^2}{2\pi K}\,(\mu-\mu_b),
\qquad
\mu(t)=\mu_b+\Delta\mu_0\,e^{-t/\tau_Q},
\qquad
\tau_Q\equiv \frac{2\pi K}{q^2},
\end{equation}
where $\Delta\mu_0\equiv \mu_0-\mu_b$. Therefore,
\begin{equation}
(\mu(t)-\mu_b)^2=\Delta\mu_0^2 e^{-\lambda t},
\qquad
\lambda\equiv \frac{2}{\tau_Q}=\frac{q^2}{\pi K},
\end{equation}
and hence,
\begin{equation}
J(t)=J_0 e^{-\lambda t},
\qquad
J_0\equiv \frac{q^2}{4\pi}\Delta\mu_0^2.
\end{equation}

To organize the expansion, it is convenient to introduce
\begin{equation}
X(t)\equiv T^2(t),
\qquad
X_b\equiv T_b^2,
\end{equation}
together with the slow time
\begin{equation}
s\equiv \gamma t,
\qquad
\gamma\equiv \frac{c}{24\pi C}.
\end{equation}
In terms of $X$, equation \eqref{A.3} becomes
\begin{equation}
X'(s)
+\frac{1}{\pi^2 C}T'(s)
=
-(X-X_b)
-\frac{3}{4\pi^2 C}(T-T_b)
-\frac{3q^2}{\pi^2 c\,K}(T-T_b)
+\frac{J(s)}{\alpha},
\label{A.10}
\end{equation}
where the prime denotes $d/ds$. We now expand the squared temperature as
\begin{equation}
X(t)
=
X_0(t)
+\frac{1}{4\pi^2 C}\,X_C(t)
+\frac{3q^2}{\pi^2 c\,K}\,X_K(t)
+\mathcal{O}\!\left(C^{-2},K^{-2},C^{-1}K^{-1}\right).
\end{equation}
The leading piece is identified with the classical solution,
\begin{equation}
X_0(t)=T_{\rm cl}^2(t).
\end{equation}
With this parametrization, the first-order temperature corrections appearing in
\eqref{A.1} are
\begin{equation}
\delta T_C(t)=\frac{X_C(t)}{8\pi^2 T_{\rm cl}(t)},
\qquad
\delta T_K(t)=\frac{3q^2}{2\pi^2 c}\,\frac{X_K(t)}{T_{\rm cl}(t)}.
\label{A.13}
\end{equation}
At leading order, equation \eqref{A.10} gives
\begin{equation}
X_0'(s)=-(X_0-X_b)+\frac{J(s)}{\alpha}.
\end{equation}
Returning to the physical time $t$, this becomes
\begin{equation}
\dot X_0+\gamma X_0=\gamma T_b^2+\delta e^{-\lambda t},
\qquad
\delta\equiv \frac{J_0}{2\pi^2 C}
=
\frac{q^2(\mu_0-\mu_b)^2}{8\pi^3 C}.
\end{equation}
For $\gamma\neq \lambda$, the solution is
\begin{equation}
X_0(t)
=
T_b^2+(T_0^2-T_b^2)e^{-\gamma t}
+\frac{\delta}{\gamma-\lambda}\left(e^{-\lambda t}-e^{-\gamma t}\right),
\label{A.16}
\end{equation}
and therefore,
\begin{equation}
T_{\rm cl}(t)=\sqrt{X_0(t)}.
\end{equation}
In the resonant case $\gamma=\lambda$, the last term in \eqref{A.16} is replaced by
\begin{equation}
\delta\, t\, e^{-\gamma t}.
\end{equation}

At first order, the $1/C$ and $1/K$ channels decouple and satisfy
\begin{equation}
X_C'(s)+X_C(s)=-4\,T_{\rm cl}'(s)-3\bigl(T_{\rm cl}(s)-T_b\bigr),
\end{equation}
and
\begin{equation}
X_K'(s)+X_K(s)=-\bigl(T_{\rm cl}(s)-T_b\bigr).
\end{equation}
Imposing the natural matching conditions
\begin{equation}
X_C(0)=0,
\qquad
X_K(0)=0,
\end{equation}
these equations integrate to
\begin{equation}
X_C(t)
=
-e^{-\gamma t}\int_0^t dt'\,e^{\gamma t'}
\left[
4\,\dot T_{\rm cl}(t')
+3\gamma\bigl(T_{\rm cl}(t')-T_b\bigr)
\right],
\label{A.22}
\end{equation}
and
\begin{equation}
X_K(t)
=
-e^{-\gamma t}\int_0^t dt'\,\gamma e^{\gamma t'}
\bigl[T_{\rm cl}(t')-T_b\bigr].
\label{A.23}
\end{equation}
Here
\begin{equation}
\dot T_{\rm cl}(t)=\frac{\dot X_0(t)}{2T_{\rm cl}(t)},
\end{equation}
with $X_0(t)$ given in \eqref{A.16}.

Substituting \eqref{A.22} and \eqref{A.23} into \eqref{A.13}, we finally obtain the
controlled first-order temperature profile
\begin{equation}
T(t)
=
T_{\rm cl}(t)
+\frac{1}{C}\,\delta T_C(t)
+\frac{1}{K}\,\delta T_K(t)
+\mathcal{O}\!\left(C^{-2},K^{-2},C^{-1}K^{-1}\right),
\end{equation}
with
\begin{equation}
\delta T_C(t)=\frac{X_C(t)}{8\pi^2 T_{\rm cl}(t)},
\qquad
\delta T_K(t)=\frac{3q^2}{2\pi^2 c}\,\frac{X_K(t)}{T_{\rm cl}(t)}.
\label{A.26}
\end{equation}
Equations \eqref{A.16}, \eqref{A.22}, \eqref{A.23}, and \eqref{A.26} are the final results of
this appendix. They provide a global controlled approximation to the quantum-corrected
temperature evolution and cleanly separate the Schwarzian $1/C$ correction from the
$U(1)$ fluctuation-induced $1/K$ correction.

For completeness, in the late-time near-equilibrium regime one may further set
\begin{equation}
T(t)=T_b+\delta T\,,
\end{equation}
which reduces the full temperature equation to the linear form used in the main text,
\begin{equation}
\dot{\delta T}=-\Gamma_T\,\delta T,
\qquad
\Gamma_T=
\frac{\pi c\,T_b/6+\Lambda}{4\pi^2 C T_b+2},
\qquad
\Lambda\equiv \frac{c}{16\pi C}+\frac{q^2}{4\pi K}.
\end{equation}

\section{Reconstruction of $\hat\eta(t)$ and $\hat F(t)$ from the controlled temperature profile}
\label{appB}
In this appendix we explain how the effective time variable $\hat\eta(t)$ and the
corresponding boundary trajectory $\hat F(t)$ are reconstructed from the controlled
temperature profile obtained from the quantum-corrected balance equation. The purpose of
this appendix is purely technical: it provides the dictionary needed to translate the dynamical
input $T(t)$ into the variables entering the entropy formulas in section \ref{sec:page-curve}.
\subsection{Exact relation between the map $\hat\eta(t)$ and $\hat F(t)$}

We introduce a vacuum coordinate $w^-$ for the outgoing modes by
\begin{equation}
w^- \equiv -e^{-2\pi \hat\eta(t)/\beta_b},
\label{B.1anch}
\end{equation}
where $\beta_b$ is the fixed inverse bath temperature. In our notation, the boundary reparametrization is denoted by $\hat F(t)$, and we choose the convenient gauge
\begin{equation}
w^-(t)=\frac{\pi \hat F(t)-\beta_b}{\pi \hat F(t)+\beta_b}.
\label{B.2anch}
\end{equation}
Combining \eqref{B.1anch} and \eqref{B.2anch}, one obtains the exact dictionary
\begin{equation}
-e^{-2\pi \hat\eta(t)/\beta_b}
=
\frac{\pi \hat F(t)-\beta_b}{\pi \hat F(t)+\beta_b},
\qquad
\hat F(t)=\frac{\beta_b}{\pi}\tanh\!\left(\frac{\pi}{\beta_b}\hat\eta(t)\right).
\label{B.3anch}
\end{equation}
At equilibrium, $\hat\eta(t)=t$, and hence
\begin{equation}
\hat F(t)=\frac{\beta_b}{\pi}\tanh\!\left(\frac{\pi t}{\beta_b}\right),
\label{B.4anch}
\end{equation}
as expected for the static thermal trajectory.

It is important that $\beta_b$, rather than the instantaneous black-hole inverse
temperature, appears in \eqref{B.1anch}--\eqref{B.3anch}. The reason is that $\beta_b$ is
the fixed thermal-cylinder parameter entering the vacuum-coordinate map, whereas the
dynamical temperature $T(t)$ enters only through the Schwarzian equation for $F(t)$.

Writing
\begin{equation}
\hat F(t)=g(\hat\eta(t)),
\qquad
g(\eta)=\frac{\beta_b}{\pi}\tanh\!\left(\frac{\pi}{\beta_b}\eta\right),
\end{equation}
and using the Schwarzian composition rule
\begin{equation}
\{g\circ \hat\eta,t\}
=
\{\hat\eta,t\}+(\hat\eta')^2\{g,\eta\},
\end{equation}
together with the standard identity
\begin{equation}
\{\tanh(a\eta),\eta\}=-2a^2,
\qquad
a=\frac{\pi}{\beta_b}=\pi T_b,
\end{equation}
we obtain
\begin{equation}
\{g,\eta\}=-2\left(\frac{\pi}{\beta_b}\right)^2=-2\pi^2T_b^2.
\end{equation}
Hence, the exact relation between $\hat F$ and $\hat\eta$ is
\begin{equation}
\{\hat F,t\}
=
\{\hat\eta,t\}-2\pi^2T_b^2(\hat\eta')^2.
\label{B.9anch}
\end{equation}
The analysis in section \ref{qcbe} determines the raw quantum-averaged Schwarzian 
\begin{equation}
\{F_{\rm raw}(t),t\}
=
-2\pi^2T(t)^2-\frac{3}{2C}T(t).
\end{equation}
If this raw Schwarzian is inserted directly into \eqref{B.9anch}, the resulting map develops
a residual linear drift even after $T(t)\to T_b$. To remove this artifact, we impose the
fixed-point normalization that the map should reduce to the standard thermal
map \eqref{B.4anch} when the system relaxes to the bath. This leads to the following definition
\begin{equation}
\{\hat F(t),t\}
:=
-2\pi^2T(t)^2-\frac{3}{2C}\bigl(T(t)-T_b\bigr).
\end{equation}
Equivalently,
\begin{equation}
\{\hat F(t),t\}
=
-2\pi^2T_b^2
-2\pi^2\bigl(T(t)^2-T_b^2\bigr)
-\frac{3}{2C}\bigl(T(t)-T_b\bigr).
\label{B.10anch2}
\end{equation}
Combining \eqref{B.9anch} and \eqref{B.10anch2}, we arrive at the exact equation
\begin{equation}
\{\hat\eta,t\}
-
2\pi^2T_b^2\Bigl[(\hat\eta')^2-1\Bigr]
=
-2\pi^2\bigl(T(t)^2-T_b^2\bigr)
-\frac{3}{2C}\bigl(T(t)-T_b\bigr).
\label{B.11anch}
\end{equation}
By construction, the right-hand side vanishes at the bath fixed point, so $\hat\eta(t)=t$
is recovered.

\subsection{Controlled adiabatic reconstruction of $\hat\eta(t)$ and $\hat F(t)$}

We now evaluate \eqref{B.11anch} in the controlled regime discussed in appendix \ref{appA}, where
we keep only first-order terms in $1/(CT)$ and $1/(KT)$, and drop
\begin{equation}
\mathcal{O}\!\left((CT)^{-2},(KT)^{-2},(CT)^{-1}(KT)^{-1}\right).
\end{equation}
In the controlled quasi-static regime, $\hat\eta'(t)$ is an $\mathcal{O}(1)$ function of the slowly
varying temperature profile $T(t)$. Since the latter evolves on the  scales $\gamma^{-1}\sim C$ and $\lambda^{-1}\sim K$, one has
\begin{equation}
\dot T = \mathcal{O}(\gamma,\lambda),\qquad
\ddot T = \mathcal{O}(\gamma^2,\lambda^2,\gamma\lambda).
\end{equation}
Therefore,
\begin{equation}
\hat\eta''(t)=\mathcal{O}(\gamma,\lambda),\qquad
\hat\eta'''(t)=\mathcal{O}(\gamma^2,\lambda^2,\gamma\lambda),
\end{equation}
and hence, from the definition of the Schwarzian derivative,
\begin{equation}
\{\hat\eta,t\}
=
\frac{\hat\eta'''}{\hat\eta'}
-\frac32\left(\frac{\hat\eta''}{\hat\eta'}\right)^2
=
\mathcal{O}(\gamma^2,\lambda^2,\gamma\lambda)
=
\mathcal{O}(C^{-2},K^{-2},C^{-1}K^{-1}).
\end{equation}

To the order of interest, this term may be neglected in \eqref{B.11anch}. We then obtain
the algebraic relation
\begin{equation}
-2\pi^2T_b^2\Bigl[(\hat\eta'(t))^2-1\Bigr]
=
-2\pi^2\bigl(T(t)^2-T_b^2\bigr)
-\frac{3}{2C}\bigl(T(t)-T_b\bigr).
\end{equation}
Equivalently,
\begin{equation}
(\hat\eta'(t))^2
=
1+\frac{T(t)^2-T_b^2}{T_b^2}
+\frac{3\bigl(T(t)-T_b\bigr)}{4\pi^2CT_b^2}.
\end{equation}
Choosing the monotone branch $\hat\eta'(t)>0$ and expanding to first order in $1/C$, we
find
\begin{equation}
\hat\eta'(t)
=
\frac{T(t)}{T_b}
+
\frac{3}{8\pi^2CT_b}\,
\frac{T(t)-T_b}{T(t)}
+\mathcal{O}(C^{-2},K^{-2},C^{-1}K^{-1}).
\end{equation}
With the gauge choice $\hat\eta(0)=0$, this integrates to
\begin{equation}
\hat\eta(t)
=
\frac{1}{T_b}\int_0^t ds\,T(s)
+
\frac{3}{8\pi^2CT_b}\int_0^t ds\,
\frac{T(s)-T_b}{T(s)}
+\mathcal{O}(C^{-2},K^{-2},C^{-1}K^{-1}).
\label{B.17anch}
\end{equation}

We now insert the controlled expansion of the temperature profile obtained in appendix \eqref{appA},
\begin{equation}
X_0(t)\equiv T_{\rm cl}(t)^2
=
T_b^2+(T_i^2-T_b^2)e^{-\gamma t}
+\frac{\delta}{\gamma-\lambda}\bigl(e^{-\lambda t}-e^{-\gamma t}\bigr),
\qquad
T_{\rm cl}(t)=\sqrt{X_0(t)},
\end{equation}
\begin{equation}
X_C(t)
=
-e^{-\gamma t}\int_0^t dt'\,e^{\gamma t'}
\Bigl[
4\,\dot T_{\rm cl}(t')+3\gamma\bigl(T_{\rm cl}(t')-T_b\bigr)
\Bigr],
\end{equation}
\begin{equation}
X_K(t)
=
-e^{-\gamma t}\int_0^t dt'\,\gamma e^{\gamma t'}
\bigl[T_{\rm cl}(t')-T_b\bigr],
\end{equation}
and
\begin{equation}
T(t)
=
T_{\rm cl}(t)
+\frac{1}{C}\,\delta T_C(t)
+\frac{1}{K}\,\delta T_K(t)
+\cdots,
\qquad
\delta T_C(t)=\frac{X_C(t)}{8\pi^2T_{\rm cl}(t)},
\qquad
\delta T_K(t)=\frac{3q^2}{2\pi^2c}\frac{X_K(t)}{T_{\rm cl}(t)}.
\label{B.21anch}
\end{equation}
Substituting \eqref{B.21anch} into \eqref{B.17anch}, we decompose
\begin{equation}
\hat\eta(t)
=
\eta_0(t)+\frac{1}{C}\,\eta_C(t)+\frac{1}{K}\,\eta_K(t)+\cdots,
\end{equation}
with
\begin{equation}
\eta_0(t)=\frac{1}{T_b}\int_0^t ds\,T_{\rm cl}(s),
\end{equation}
\begin{equation}
\eta_C(t)
=
\frac{1}{T_b}\int_0^t ds\,\delta T_C(s)
+
\frac{3}{8\pi^2T_b}\int_0^t ds\,
\frac{T_{\rm cl}(s)-T_b}{T_{\rm cl}(s)}
=
\frac{1}{8\pi^2T_b}\int_0^t ds\,
\frac{X_C(s)+3\bigl(T_{\rm cl}(s)-T_b\bigr)}{T_{\rm cl}(s)},
\label{B.28}
\end{equation}
and
\begin{equation}
\eta_K(t)
=
\frac{1}{T_b}\int_0^t ds\,\delta T_K(s)
=
\frac{3q^2}{2\pi^2 c\,T_b}\int_0^t ds\,
\frac{X_K(s)}{T_{\rm cl}(s)}.
\label{B.29}
\end{equation}
Finally, using the exact relation \eqref{B.3anch} with $\hat\eta$, we obtain
\begin{equation}
\hat F(t)
=
\frac{\beta_b}{\pi}\tanh\!\left(\frac{\pi}{\beta_b}\hat\eta(t)\right).
\end{equation}
If one expands around the zeroth-order trajectory, define
\begin{equation}
\delta\eta(t)\equiv \hat\eta(t)-\eta_0(t),
\qquad
F_0(t)\equiv \frac{\beta_b}{\pi}\tanh\!\left(\frac{\pi}{\beta_b}\eta_0(t)\right),
\end{equation}
then to first order,
\begin{equation}
\hat F(t)\simeq
F_0(t)+\delta\hat\eta(t)\,
{\rm sech}^2\!\left(\frac{\pi}{\beta_b}\eta_0(t)\right).
\end{equation}

\bibliographystyle{JHEP}
\bibliography{rose}

\end{document}